\documentclass[journal]{IEEEtran}
\pdfoutput=1
\IEEEoverridecommandlockouts
\usepackage{cite}
\usepackage{amsmath,amssymb,amsfonts}
\usepackage{bm}
\usepackage{mathrsfs}
\usepackage{multirow}
\usepackage{amsfonts,amssymb}
\usepackage{tabu}
\usepackage[colorlinks, linkcolor=black, citecolor=black, urlcolor=black]{hyperref}

\usepackage{subfigure}
\usepackage{graphicx}
\usepackage{textcomp}
\usepackage{xcolor}
\def\BibTeX{{\rm B\kern-.05em{\sc i\kern-.025em b}\kern-.08em
    T\kern-.1667em\lower.7ex\hbox{E}\kern-.125emX}}
\usepackage{float}
\usepackage{hyperref}
\usepackage[hyphenbreaks]{breakurl}
\usepackage{booktabs,multirow,array,longtable}
\usepackage{float}
\newcommand{\tabincell}[2]{\begin{tabular}{@{}#1@{}}#2\end{tabular}}
\usepackage{algorithm,algorithmic}

\usepackage{xparse}
\usepackage{blindtext}
\usepackage{morewrites}
\usepackage{wrapfig}
\usepackage{xkeyval}%

\begin{document}


\title{Machine Learning based Malicious Payload Identification in Software-Defined Networking}

\author{Qiumei Cheng,~\IEEEmembership{Student Member,~IEEE,}
        Chunming Wu,~\IEEEmembership{Member,~IEEE,}
        Haifeng Zhou,~\IEEEmembership{Member,~IEEE,}
        Dezhang Kong,
        Dong Zhang,~\IEEEmembership{Member,~IEEE,}
        Junchi Xing,
        Wei Ruan,
\thanks{\textit{Corresponding author: Chunming Wu. (email: wuchunming@zju.edu.cn)}}
}

\maketitle

\begin{abstract}
Deep packet inspection (DPI) has been extensively investigated in software-defined networking (SDN) as complicated attacks may intractably inject malicious payloads in the packets. Existing proprietary pattern-based or port-based third-party DPI tools can suffer from limitations in efficiently processing a large volume of data traffic. In this paper, a novel OpenFlow-enabled deep packet inspection (OFDPI) approach is proposed based on the SDN paradigm to provide adaptive and efficient packet inspection. First, OFDPI prescribes an early detection at the flow-level granularity by checking the IP addresses of each new flow via OpenFlow protocols. Then, OFDPI allows for deep packet inspection at the packet-level granularity: (i) for unencrypted packets, OFDPI extracts the features of accessible payloads, including tri-gram frequency based on Term Frequency and Inverted Document Frequency (TF-IDF) and linguistic features. These features are concatenated into a sparse matrix representation and are then applied to train a binary classifier with logistic regression rather than matching with specific pattern combinations. In order to balance the detection accuracy and performance bottleneck of the SDN controller, OFDPI introduces an adaptive packet sampling window based on the linear prediction; and (ii) for encrypted packets, OFDPI extracts notable features of packets and then trains a binary classifier with a decision tree, instead of decrypting the encrypted traffic to weaken user privacy. A prototype of OFDPI is implemented on the Ryu SDN controller and the Mininet platform. The performance and the overhead of the proposed sulotion are assessed using the real-world datasets through experiments. The numerical results indicate that OFDPI can provide a significant improvement in detection accuracy with acceptable overheads.
\end{abstract}




\section{Introduction}
Recently, software-defined networking (SDN) has been considered as a popular paradigm that separates the control logic from underlying forwarding switches in a centralized manner, to enable fine-grained flow management \cite{Mckeown2008OpenFlow}. Meanwhile, soaring traffic traversing SDN-based network has led to potential security issues for network administrators. The attacks, e.g., the malicious packet payload injection, can directly lead to serious consequences, e.g., cross-site scripting and SQL injection, which are responsible for just over 51$\%$ of web attacks \cite{XXS}. SDN provides flow-level verification to monitor layer1-layer4 traffic\cite{Javid2014A, Suh2014Building, Kaur2015Programmable, Othman2017} through matching against packet header fields with tuple space search \cite{Srinivasan1999Packet}. However, the flow-level features are generally unreliable for in-depth payload inspection to monitor the actual packet contents without headers, and hence it can be circumvented by attackers.

To identify the high-level anomaly traffic, the researchers explore to provide deep packet inspection (DPI) service in the SDN paradigm. Existing DPI schemes in SDN can be divided into two categories. First, SDN has naturally resorted to third-party deep packet inspection (DPI) tools, e.g., OpenDPI \cite{Bela2011}, nDPI \cite{nDPI}, and L7-filter \cite{Crotti2007Traffic}, accomplished by incorporating DPI modules into specific vendor proprietary hardware or transforming DPI into middleboxes. Specifically, network function virtualization (NFV) opens up new venues for DPI approaches by mirroring flows to specific modules. In particular, \cite{Huang2017Traffic, Bouet2013Cost, lin2015} extend the SDN architecture to redirect traffic to the specified DPI proxies. Apart from  traffic redirection to DPI proxies, some researches \cite{ErikLarsson,Cho2016A} turn their eyes on developing DPI modules in OpenFlow switch (OvS) to avoid additional routing hop. Nevertheless, enabling application-aware functionality in the data plane will result in dramatic performance degeneration in network nodes, whose QoS demands can no longer be guaranteed \cite{Li2017Deep}.  In fact, whether implementing DPI modules in the data plane or in a remote proxy, these approaches usually utilize open-source DPI tools that are largely pattern-based or port-based \cite{Finsterbusch2014A}.  Even though pattern matching is valid for checking payloads of captured packets, this approach still remains limitations in efficiently processing large volume of traffic \cite{Lin2008Using}. With the recognition of  exponential growth of network traffic, packet inspection with high throughput as well as efficient intrusion detection is demanded.

The second category gains inspiration from machine learning and deep learning technologies. Recently, machine learning algorithms have emerged as a promising solution to identify anomalous packets, which are widely explored in many works \cite{Wang2016, Cusack2018, Zhang2013Unsupervised}. The payloads inside packets are sometimes text strings indeed (e.g., web queries), and thus such packet metadata can provide high-level information. Since the feature extraction in traditional machine learning is complicated and time-consuming, more and more researches pay attention to provide end-to-end detection with deep learning. \cite{2019Dela} proposes a conceptual SDN-based security monitoring framework for the smart grid with behavioral analysis and deep learning models. \cite{CNN2018} provides an end-to-end payload classification without feature extraction using CNN-based and RNN-based classification approaches to learn features from raw payloads. Even though such machine learning or deep learning approaches achieve high detection accuracy, they can only handle unencrypted payloads.

Nevertheless, network traffic is increasingly transmitted with encryption by using the Transport Layer Security (TLS) protocol, covering 80$\%$ web traffic in 2019, estimated by Gartner \cite{Gartner2016}. Meanwhile, the attackers will use encryption to conceal malware traffic. In light of encrypted traffic, a typical approach is to use decryption to obtain the payloads of the packets, which unfortunately weakens the user privacy. BlindBox \cite{Sherry2015} claims the first attempt to enable deep packet inspection on encrypted traffic without decrypting the payloads. Apart from supporting keyword matching and regular expression like \cite{Sherry2015}, SPABox \cite{Fan2017} also supports malware detection via machine learning with a limited connection setup overhead. Similarly, \cite{Anderson2016Identifying} safeguards user privacy by developing supervised machine learning models with contextual data. Machine learning techniques shed light on the deep packet inspection in SDN for both unencrypted payloads and encrypted payloads.

However, the integration of machine learning methods and the SDN paradigm still inherits some challenges. Specifically, how to customize the SDN controller with a lower overhead when processing large volume of traffic? Inspecting all the packets in the SDN paradigm brings a heavy burden to the SDN controller \cite{Shah2013An}. The SDN controller needs to deal with the huge number of OpenFlow (OF) messages, at the expense of frequent redundant memory copy, incurring high CPU utilization. The imperfection of the current DPI schemes necessitates the development of an efficient deep packet inspection approach in SDN \cite{Chin2015}.

To this end, this paper proposes a novel OpenFlow-enabled deep packet inspection (OFDPI) approach in the software-defined networking paradigm by utilizing machine learning techniques. The main idea behind the proposed approach is as follows. OFDPI extracts the features of payloads and then train a binary logistic regression classifier. The payload features include tri-gram frequency based on Term Frequency and Inverted Document Frequency (TF-IDF) and linguistic features, e.g., the number of digits in the payload, the number of consecutive consonants. To seek a trade-off between detection accuracy and resource constraints of Ryu SDN controller, OFDPI devises a packet sampling window based on the linear prediction. Prior to in-depth payload inspection, OFDPI prescribes an early detection at flow-level by checking the IP addresses of each new flow via OpenFlow protocols, with an adoption of a pre-defined IP blacklist from Cisco. Besides, OFDPI also allows for the inspection of encrypted packets. Instead of decrypting the encrypted traffic to weaken user privacy, OFDPI extracts some notable features that can distinguish between malicious encrypted packets and benign packets to provide a statistical machine learning solution.

In this work, a prototype of OFDPI is further implemented on the Ryu SDN controller and the Mininet platform. Two real-world datasets are used to train the binary classifiers for unencrypted traffic and encrypted traffic, respectively. For unencrypted packets inspection, a binary logistic regression classifier outperforms other typical classifiers with a rather high detection accuracy of 98.86$\%$. For encrypted packets inspection, a decision tree classifier achieves a high accuracy of 99.15$\%$. To avoid overfitting with a particular dataset, another unencrypted dataset is adopted to evaluate the performance of the binary logistic regression classifier. The main technical contributions made in this paper are manifold:
\begin{itemize}
    \item This paper proposes a novel OpenFlow-enabled deep packet inspection (OFDPI) approach in the SDN paradigm incorporating with machine learning algorithms.  Prior to deep packet inspection, OFDPI prescribes an early detection at flow-level granularity by checking the IP addresses of each new flow via OpenFlow protocols.
    \item For unencrypted packets, OFDPI extracts the payload features (TF-IDF and linguistic features) and then trains a binary logistic regression classifier to perform packet inspection.  To reduce the resource constraints of the SDN controller, OFDPI devises a packet window based on linear prediction and then performs adaptive packet sampling.
    \item For encrypted packets, OFDPI extracts some notable features of encrypted traffic (e.g., TLS ciphersuites) to train a decision tree classifier in identifying malicious encrypted packets, rather than decrypt the encrypted traffic to weaken user privacy.
    \item A prototype of OFDPI is implemented under the Ryu SDN controller and the Mininet platform to evaluate the performance and overheads of OFDPI. The experiments results show that OFDPI can provide high detection accuracy with acceptable overheads.
\end{itemize}

The rest of this paper is organized as follows. Section~\ref{related work} presents the related work. Section~\ref{overview} briefly introduces the overview of OFDPI. This is followed by an elaborate design of OFDPI in Section~\ref{design}. The evaluation of the proposed scheme in section~\ref{evaluation} covers an offline training test and online analysis. Section~\ref{conclusion} makes the concluding remarks of this paper.

\section{Related Work}
\label{related work}
An abundance of literature has investigated the issue of deep packet inspection in the SDN paradigm. Generally, previous researches have been predominant to consider the deployment of DPI engines on proprietary hardware to seek a trade-off between the number of DPI engines and network resource constraints.

Huang et al. \cite{Huang2017Traffic} provide a deep packet inspection method for traffic traversing the ingress switches in the context of an integrated DPI proxy allocation and routing determining problem. They formulate it as an ILP problem to minimize the overall delay of DPI processing. Accordingly, a two-phase heuristic algorithm is proposed to select the DPI proxy and determine corresponding routing paths. For better routing strategy decision, \cite{Adami2015} enables a DPI module in the control application to detect Session Initial Protocol (SIP) signaling messages, and thus applying a differentiated routing strategy. Bouet et al. \cite{Bouet2013Cost} optimize the deployment of DPI engines in SDN environment from a cost perspective. Considering operational constraints regarding maximum bandwidth per link and license fees of DPI engine, they minimize the number of DPI engines with genetic algorithms to reach a trade-off, reduce up to 58$\%$ of global cost. To reduce the overhead of the controller, \cite{lin2015} extends the conventional SDN architecture with a two-tier traffic classification mechanism. First, it performs traffic classification on the data plane rather than the controller. If the first classification module is unable to handle traffic, the traffic will be redirected to a DPI module. However, redirecting traffic to particular DPI engines inevitably leads to global network load and link utilization issues.

Packet processing delay and redundant memory copy motivate researchers to turn their eyes on developing DPI modules in OpenFlow switch to avoid additional routing hop, e.g., a pre-integrated DPI engine embedded in the OpenvSwitch (OvS), analogous to a next-generation network gateway. The DPI placed at the data link layer may sound applicable for efficient packet forwarding at line rate, but it should be noted that it partially violates the principle of SDN, in which OpenFlow switches are only capable of dispatching packets without any built-in application-aware intelligent logic. ChoongHee Cho et al. \cite{Cho2016A} pay attention to payload information in the packets rather than only utilizing header information for packet classification. They propose a sophisticated packet forwarding scheme by implementing a DPI module in a virtual OpenFlow switch associated with a monitoring application. If the DPI module detects a pre-defined string pattern in a packet, the OpenFlow switch will send information to a log server where the monitoring application periodically gathers log information. Nevertheless, they only utilize a pre-defined string pattern based on Snort. Li et al. \cite{Li2017Deep} introduce DPI into application-aware traffic control plane of SDN with extensions of flow table, but they only send the first packet to the SDN controller for identification in consideration of the performance of the SDN controller. Actually, implementing the DPI in the data plane or in remote proxy is largely pattern-based or port-based.

Recently, machine learning and deep learning methods have been applied to develop classifiers to analyze payloads rather than traditional pattern-based models. Wang et al. \cite{Wang2016} classify the network traffic according to the QoS requirement in SDN. However, the traffic classification engine implemented in the SDN controller may cause the bottleneck of the controller. \cite{Cusack2018} collects the traffic with programmable forwarding engines (PFEs) and then detects ransomware by utilizing a random forest binary classifier. Since the feature extraction in traditional machine learning is complicated and time-consuming, more and more researches pay attention to provide end-to-end detection with deep learning. \cite{2019Dela} proposed a conceptual SDN-based security monitoring framework for the smart grid, providing cyber-attacks detection with behavioral analysis and deep learning models. Unfortunately, the conceptual framework was shorted of experimental validation. Liu et al. \cite{CNN2018} present a convolutional neural network-based (CNN) and a recurrent neural network-based (RNN) approach in attack detection, performed by learning features from payloads in the packets. Even though \cite{CNN2018} achieves a high detection accuracy, it is unable to handle encrypted payloads. BlindBox \cite{Sherry2015} claims the first attempt to enable deep packet inspection on encrypted traffic without decrypting the payloads. Apart from supporting keyword matching and regular expression like \cite{Sherry2015}, SPABox \cite{Fan2017} also supports malware detection via machine learning with a limited connection setup overhead. Similarly, \cite{Anderson2016Identifying} safeguards user privacy by developing supervised machine learning models with contextual data. Machine learning techniques shed light on the deep packet inspection in SDN for both unencrypted payloads and encrypted payloads.

In this paper, we implement an OFDPI scheme in SDN by utilizing machine learning techniques to perform deep packet inspection for both unencrypted packets and encrypted packets. Our earlier work explored the payload inspection in SDN with machine learning techniques \cite{Cheng2018}. However, the caused overhead is out of consideration. Rather than identifying the first packet of a flow in \cite{Li2017Deep}, we aim to sample partial packets for deep packet inspection to maintain a trade-off of the detection accuracy and the performance of the controller. We get inspirations from adaptive traffic sampling according to linear prediction\cite{Hernandez2001Adaptive, Zhang2013An, Silva2017Inside} to avoid redundant memory copy and long processing time. Some works implement a linear prediction-based algorithm to balance the monitoring overhead and anomaly detection accuracy \cite{Zhang2013An}, but they always conduct a time-driven sampling model. Instead, we employ a packet-driven linear prediction model based on an adaptive packet window to enable packet sampling according to previous samples.

\section{Overview}
\label{overview}
\begin{figure*}[htb]
	\centering
	\includegraphics[width=12cm]{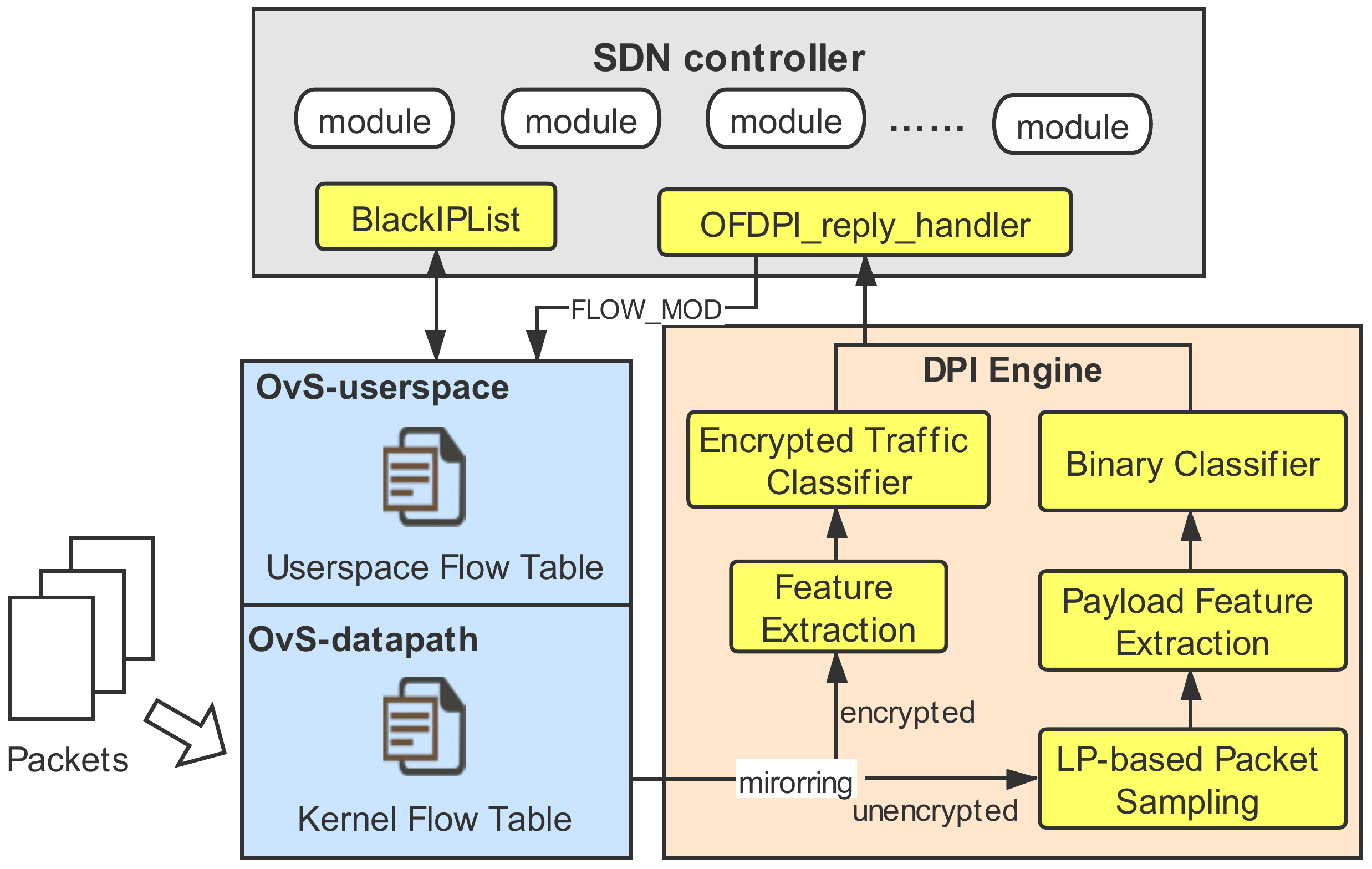}
	\caption{The solution overview}
	\label{Fig:framework}	
\end{figure*}

In this section, we briefly introduce the overview of the proposed OFDPI approach, as illustrated in Figure~\ref{Fig:framework}. OFDPI enables deep packet inspection for both unencrypted traffic and encrypted traffic.

OFDPI allows for a two-stage deep packet inspection mechanism. Prior to deep packet inspection, OFDPI prescribes an early detection at flow-level granularity, accomplished by checking the IP addresses of each new flow via OpenFlow protocols with a pre-defined IP blacklist from Cisco. For those packets that match against flow entries in the kernel, in-depth payload analysis is necessary in case malicious packets evade the early detection phase.

In the second stage,  the packets will be redirected to a DPI engine located in another server within the same datacenter by port mirroring,  as illustrated in Figure~\ref{Fig:framework}. The DPI engine is associated with two parts: unencrypted packets inspection and encrypted packets inspection. In order to reduce the overhead of inspection, we implement an adaptive traffic sampling mechanism in the DPI engine, in which the packet window based on linear prediction performs packet sampling by analyzing the previous consecutive packets. An $OFDPI\_reply\_handler$ residing in the SDN controller is in charge of receiving the notification of the DPI engine and determining the stay of the packets. If the packet is confirmed anomalous, an \texttt{FLOW\_MOD} message will be installed to block traffic accordingly.

For \emph{unencrypted} packets, the payloads of the packets are easily accessible to extract features and then perform in-depth packet analysis. The payload features include tri-gram frequency based on Term Frequency and Inverted Document Frequency (TF-IDF) and linguistic features, e.g., the number of digits in the payload, the number of consecutive consonants. These features are concatenated into a sparse matrix representation and are then applied to train a binary classifier since there are two types of traffic: anomalous and benign. A binary classifier is training offline since it is generally time-consuming in machine learning, which is susceptible to real-time traffic transmission. However, analyzing all the payloads brings a huge cost to the system. To seek a trade-off between detection accuracy and resource constraints of the SDN controller, a packet window based on linear prediction is devised to perform adaptive packet sampling by analyzing some consecutive packets.

For \emph{encrypted} packets, payloads are invisible to the network administrators. Rather than decrypt network traffic, this paper provides statistical machine learning solutions by employing some traffic features. After investigating a real encrypted botnet dataset, some notable features can distinguish between malicious traffic and benign traffic, including TTL, TLS Ciphersuites, duration, ports, and etc. OFDPI extracts these distinguished features to train a binary classifier under supervised learning. The online classification results provide information for remediation and also periodically updates offline training dataset.
\section{OFDPI}
\label{design}
Given a hierarchically structured overview of OFDPI, this section presents the design rationale of OFDPI in detail. First, OFDPI introduces an early detection mechanism associated with IP filtering at flow-level. Then, the packets are monitored by a DPI engine located in another server, which consists of two main components: unencrypted traffic inspection and encrypted traffic inspection. Talbe. \ref{tab:notations} lists the key notations of this paper.
\begin{table}[htb]
  \centering
  \scriptsize
  \caption{Notations used in mathematical models}
  \label{tab:notations}
  \begin{tabular}{ll}
    \\[-0mm]
    \hline
    \hline\\[-2mm]
    {\bf \small Symbol}&\qquad {\bf\small Description}\\
    \hline
    \vspace{1mm}\\[-3mm]
    $\delta _{i}$      &   \tabincell{l}{The real number of malicious packets in the $i\,th$ sample}\\
    \vspace{1mm}
    $w_{i}$          &  \tabincell{l}{ The window size of the $i\,th$ sampling}\\
     \vspace{1mm}
    $m$          &  \tabincell{l}{constant, $m\gg w_{i}, \forall i \in R$}\\
     \vspace{1mm}
    $\hat{\delta }_{n+1}$  &   \tabincell{l}{The predicted malicious packets in the $\left ( n+1 \right )-th$ sample }\\
     \vspace{1mm}
    $\Delta w$    & \tabincell{l}{The variation of the packet window size between $w_{n+1}$ and $w_{n}$}\\
     \vspace{1mm}
     $\Delta w^{'}$  &   \tabincell{l}{The variation of the packet window size in the next sample }\\
     \vspace{1mm}
     $W_{min}$  &   \tabincell{l}{The minimun value of the packet window size}\\
          \vspace{1mm}
     $W_{max}$  &   \tabincell{l}{The maximum value of the packet window size}\\
               \vspace{1mm}
     $l_{i}^{j}$  &   \tabincell{l}{The actual value of $j\,th$ linguistic feature in the $i\,th$ payload}\\
                    \vspace{1mm}
      $N_{i}^{j}$  &   \tabincell{l}{The normalized value of $l_{i}^{j}$}\\
                          \vspace{1mm}
      $l_{min}$  &   \tabincell{l}{The minimum value of the linguistic features}\\
                                \vspace{1mm}
      $l_{max}$  &   \tabincell{l}{The maximum value of the linguistic features}\\

    \hline
    \hline
  \end{tabular}
\end{table}

\subsection{Early Detection}
Before in-depth payload inspection, OFDPI maintains a simple early detection mechanism with the consideration of that IP address filtering is still necessary especially for elephant flow, for which partial payload inspection is not reliable.
\begin{figure}[htb]
	\centering
	\includegraphics[width=9cm]{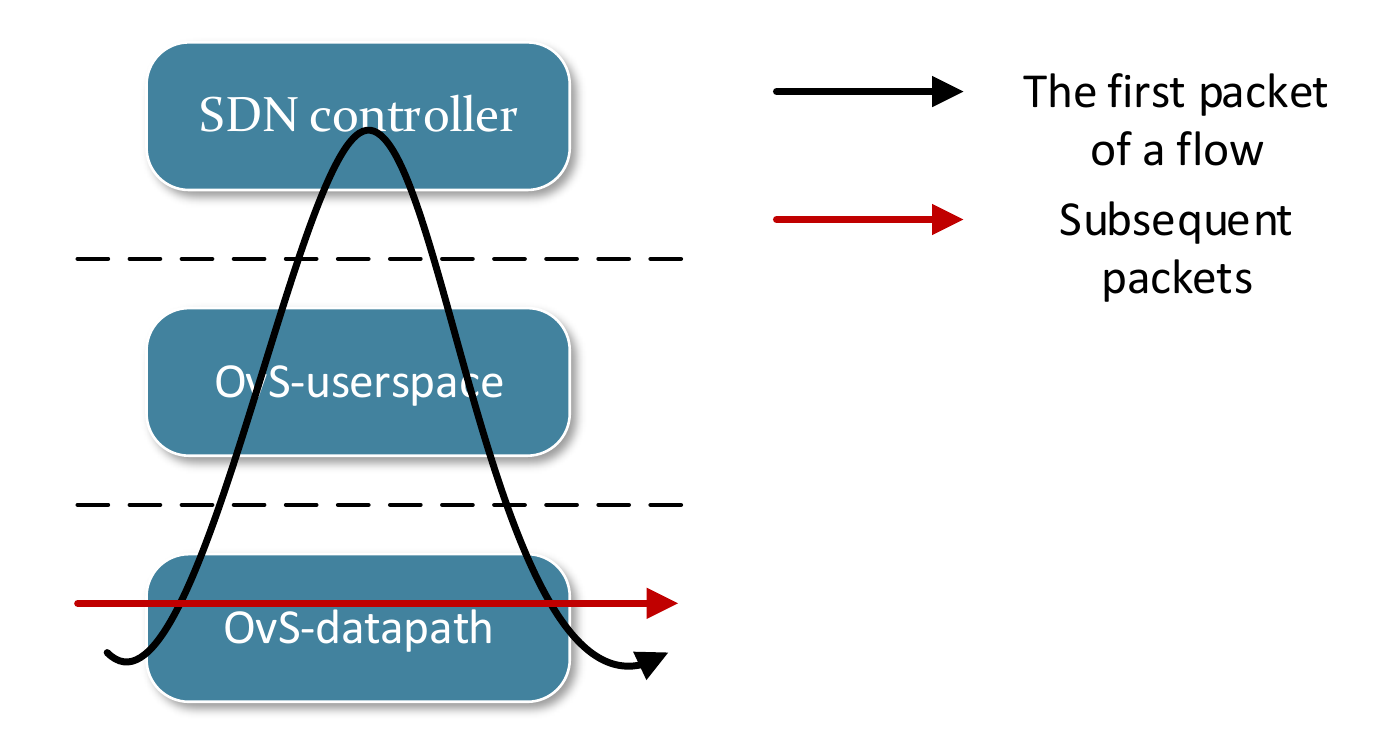}
	\caption{Workflow of packet processing in SDN}
	\label{fig:sdnworkflow}	
\end{figure}

\textbf{Definition 1 (Traffic Flow)} \emph{A traffic flow $F$ is a bi-directional communication between the source node and the destination node. A flow is composed of a sequence of packets. Each packet is associated with a 5-tuple: ( sIP, sPort, dIP, dPort, pro ), which represents the source IP address, source port, destination IP address, destination port and protocol type, respectively.}

Fig.\ref{fig:sdnworkflow} illustrates the workflow of packet processing in SDN. All packets from the external network first flow through datapath module in the OvS kernel space, and then OvS will extract key values (e.g., MAC layer message, network layer message) to match against flow tables cached in the kernel \cite{Pfaff2015The}. If a match fails (for the first packet in a flow), the datapath module will append the packet to trigger \emph{upcall} instead, a method that notifies the OvS user space (i.e., OvS-vSwitchd module) where a packet is received. For packets that do not have a matching flow entry in user space, a \texttt{Packet\_in} message is always generated to inquire the controller. It is unnecessary for the subsequent packets to generate \texttt{Packet-in} message if they successfully match against flow tables cached in the kernel.

To perform early detection, a blacklist of IP addresses is available in the SDN controller to monitor the source IP address of the packet in this paper. Once receiving the first packet, the SDN controller decodes it and extracts key values of network layer message. If the SDN controller verifies the source IP address of the packet within the flow is illegal, the SDN controller blocks this flow via OpenFlow messages accordingly. Otherwise, the SDN controller installs corresponding flow entries in the OvS. Deep packet inspection is necessary only if the former packet succeeds in a flow table lookup.

\subsection{Unencrypted Packets Inspection}
For unencrypted traffic, payloads of the packets are accessible for deep packet inspection. However, analyzing all the payloads brings a huge cost to the system. To this end, this paper devises an adaptive packet-driven sampling mechanism based on linear prediction to address the dilemma of resources and performances in packet-level granularity.

\subsubsection{LP-based Adaptive Packet Sampling}
\label{LPsample}
First, a packet sampling window is defined in the following.

\textbf{Definition 2 (Packet Sampling Window)}\emph{A packet sampling window captures a collection of consecutive packets selected for deep packet inspection. The packet window size equals the number of the sampled packets.}

\begin{figure*}[htb]
	\centering
	\includegraphics[width=12cm]{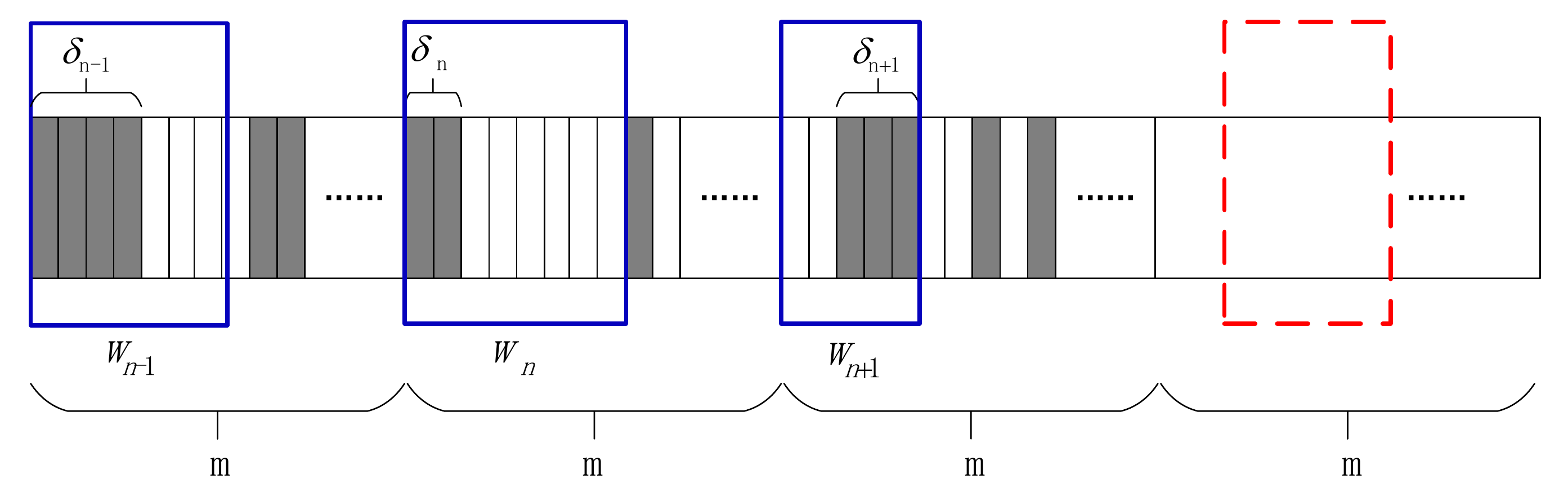}
	\caption{Adaptive packet sampling with the scalable packet window}
	\label{Fig:packetwindow}	
\end{figure*}
Fig.\ref{Fig:packetwindow} depicts the scalable packet sampling window mechanism. The central principle of LP-based adaptive packet sampling attempts to estimate the packet window size of the next sample to capture consecutive $w_{i}$ packets from every $m$ packets of a flow, where $m\gg w, \forall i \in R$. $m$ is a constant which is experimentally determined. The predicted sampled packets $\hat{\delta }_{n+1}$ in the (n+1)-th sample is written as a linear combination of the previous $n$ samples:
\begin{equation}
\hat{\delta }_{n+1}=\delta _{n}+\frac{\Delta w}{n-1}\sum_{i=1}^{n-1}\left ( \frac{\delta _{i+1}-\delta _{i}}{w_{i+1}-w_{i}} \right ), n\geq 2
\label{Eq:Yp}
\end{equation}
where $\delta_{i}$ denotes the real number of malicious packets in the i-th sample, and $\Delta w$ represents the current variation of the packet window size between $w_{n+1}$ and $w_{n}$ and is calculated as:
\begin{equation}
\Delta w=w_{n+1}-w_{n}
\label{Eq:deltaW}
\end{equation}
$\Delta w$ can be either positive or negative. If $\Delta w > 0$, the current packet window size $w_{n+1}$ is larger than the previous packet window size. If $\Delta w < 0$, it indicates that $w_{n+1}$ is smaller than the previous packet window size. If $\Delta w = 0$, it demonstrates that the packet window size remains unchanged.

This paper aims to control the variation of the packet window size in the next sample, $\Delta w^{'}$. A set of rules are established to adjust $\Delta w^{'}$ by analyzing the real number of malicious packets and the predicted value. According to the rate of change in the prediction and the rate of change in the real samples, we obtain that:
\begin{equation}
\frac{\hat{\delta }_{n+1}-\delta_{n}}{\Delta w^{'}}=\frac{\delta_{n+1} -\delta_{n}}{\Delta w}
\end{equation}
If we use a ratio metric $R\left ( \hat{\delta }_{n+1} \right )=\frac{\Delta w^{'}}{\Delta w}$, then the ratio $R\left ( \hat{\delta }_{n+1} \right )$ can be written as:
\begin{equation}
R\left ( \hat{\delta }_{n+1} \right )=\frac{\hat{\delta }_{n+1}-\delta_{n}}{\delta_{n+1} -\delta_{n}}
\end{equation}

If $\delta_{n+1} = \delta_{n}$, $R\left ( \hat{\delta }_{n+1} \right )$ is indeed undefined and thus the $\Delta w^{'}$ can be computed as:
\begin{equation}
\Delta w^{'}=\begin{cases}
5 & \text{ if } w_{n+1}= w_{n}\\
-\frac{1}{2}\times\Delta w  & \text{ otherwise}
\end{cases}
\label{Wnext}
\end{equation}
Note that when $w_{n+1}= w_{n}$, this case indicates that the number of sampled malicious packets in two samples with the same packet window size is the same. In this case, it requires an enlarged packet window size (e.g., set $\Delta w^{'} = 5$) to determine the new traffic pattern. When $w_{n+1}> w_{n}$, it indicates that the packet window should be narrow down and vice versa.

The variation of the packet window size in the next sample $\Delta w^{'}$ on most occasions is given by Eq.~\ref{Eq:Wnext}.
\begin{equation}
\Delta w^{'}=-\frac{\hat{\delta }_{n+1}-\delta_{n+1} }{\left | \hat{\delta }_{n+1}-\delta_{n+1}  \right |}\cdot \left | R\left (\hat{\delta }_{n+1} \right )\cdot \Delta w \right |
\label{Eq:Wnext}
\end{equation}
If $\hat{\delta }_{n+1}=\delta_{n+1} $, let $\Delta w^{'}=0$ which keeps the same packet window size as $w_{n+1}$. Therefore, we are able to obtain the packet window size of next sample, $w_{n+2}=w_{n+1}+\Delta w^{'}$. Considering the storage of the measured data is limited, larger window size requires much more processing resources correspondingly. The range of values satisfies a condition: $W_{min}\leq w_{n+2}\leq W_{max}$, where $W_{min}$ denotes the minimum value of the packet window size and the $W_{max}$ is the maximum value of the packet window that cannot be exceeded.

\begin{algorithm}[ht]
	
	\caption{ Adaptive Packet Sampling  }
	\begin{algorithmic}[1]
		\label{samplingpayload}
		\REQUIRE
		the packet window size in previous $n+1$ samples $\left [ w_{1}, w_{2}, \cdots, w_{n+1}\right] $; malicious packets in previous $n+1$ samples $\left [ \delta_{1}, \delta_{2}, \cdots, \delta_{n+1}\right] $ ; the minimum and maximun packet window size $W_{min}$, $W_{max}$;

		\STATE The current variation of the packet window size $\Delta w=w_{n+1}-w_{n}$;
		\STATE Calculate the predicted sampled packets in the $n+1$ sample: $\hat{\delta }_{n+1}=\delta _{n}+\frac{\Delta w}{n-1}\sum_{i=1}^{n-1}\left ( \frac{\delta _{i+1}-\delta _{i}}{w_{i+1}-w_{i}} \right ), n\geq 2$;
		
		\IF {$\delta_{n+1} = \delta_{n}$}
		\STATE Compute the variation of the packet window size in the next sample $\Delta w^{'}$ according to Eq.(5);
		\ELSE
		\STATE Calculate a ratio metric $R\left ( \hat{\delta }_{n+1} \right )=\frac{\hat{\delta }_{n+1}-\delta_{n}}{\delta_{n+1} -\delta_{n}}$ according to the rate of change in the prediction and the rate of change in the real samples;
		\STATE Compute the variation of the packet window size in the next sample $\Delta w^{'}=-\frac{\hat{\delta }_{n+1}-\delta_{n+1} }{\left | \hat{\delta }_{n+1}-\delta_{n+1}  \right |}\cdot \left | R\left (\hat{\delta }_{n+1} \right )\cdot \Delta w \right |$;
		\ENDIF
		\STATE Adjusting the packet window size of the next sample $w_{n+2}=w_{n+1}+\Delta w^{'}$ $s.t.$ $W_{min}\leq w_{n+2}\leq W_{max}$;
		
	\end{algorithmic}
	
\end{algorithm}

The adaptive packet sampling algorithm based on linear prediction is demonstrated in Algorithm ~\ref{samplingpayload}. On the observation of previous samples, the LP-based packet sampling mechanism attempts to estimate the packet window size of the next sample. The variation between the predicted value and the actual value helps the network administrator flexibly and timely adjust the packet window size, providing an adaptive packet sampling in a proactive way.

\subsubsection{Feature Extraction}
\label{sec:feature}
Payloads are contents of the packets without packet headers, which are the actual data inside of a packet. The payloads within the packets are sometimes text strings associated with characters (e.g., web queries). A \emph{payload} is a sequence of characters which consists of numbers and alphabets. However, the machine learning algorithm is incapable of directly processing the payloads in the string format. It is necessary to create the payload representation that reflects semantic relationships. To address this issue, word embedding that maps a word to a vector with a dictionary is utilized to obtain data features for further modeling. We define a feature vector of each payload within the packets, which concatenates all the extracted features, including term frequency and linguistic features.

\textbf{TF-IDF}.
Generally, some words appear in higher frequency in all the payloads, e.g., 'jsp,' php,' 'asp,' 'script,' etc., which are of little relevance to a specific payload. Term Frequency-Inverse Document Frequency (TF-IDF) tries to allocate such frequent words with lower weights but increase the importance of those words significant to a particular payload, which reflects how important a specific word to a payload, and it can be written as:
\begin{equation}
TF-IDF = TF * IDF
\end{equation}
where $TF$ represents the frequency of a specific word in a given payload. The word with low frequency in a document often results in bigger $IDF$, which demonstrates that this word is much better for payload classification. TF-IDF aims to get numerical features of payload strings rather than directly handle encoded payload data, in particular, obtaining feature vectors from the payload.

In order to calculate the value of TF-IDF, trigrams is exploited to segment payloads. A tri-gram of the payload is a set of all character sequences with the length of 3. For example, if a payload is "/javascript/debug.exe", the payload will be divided into several segmentations, like '/ja', 'jav', 'ava', 'vas', 'asc', 'scr', 'cri', 'rip', 'ipt', 'pt/', 't/d', '/de', 'deb', 'ebu', 'bug', 'ug.', 'g.e', '.ex', 'exe'.

\begin{figure*}[htbp]
  \centering
  \subfigure[Illustration of the anomalous payloads]{
    \label{fig:badcloud} 
    \includegraphics[width=5.5cm]{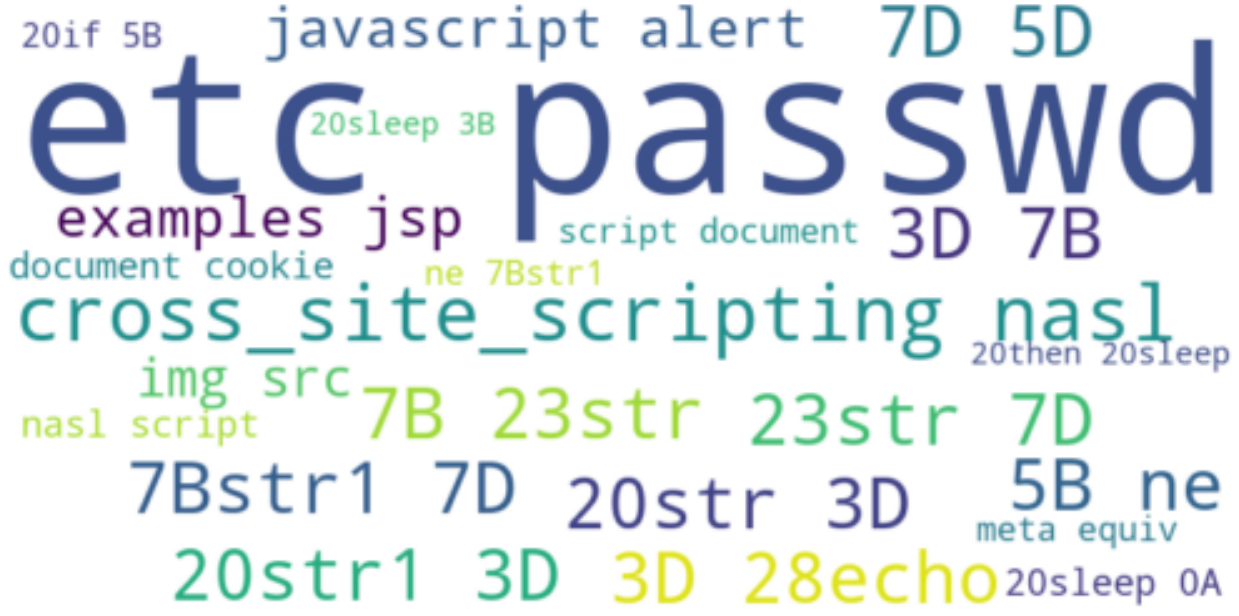}
  }
  \subfigure[Illustration of the normal payloads]{
    \label{fig:goodcloud} 
    \includegraphics[width=5.5cm]{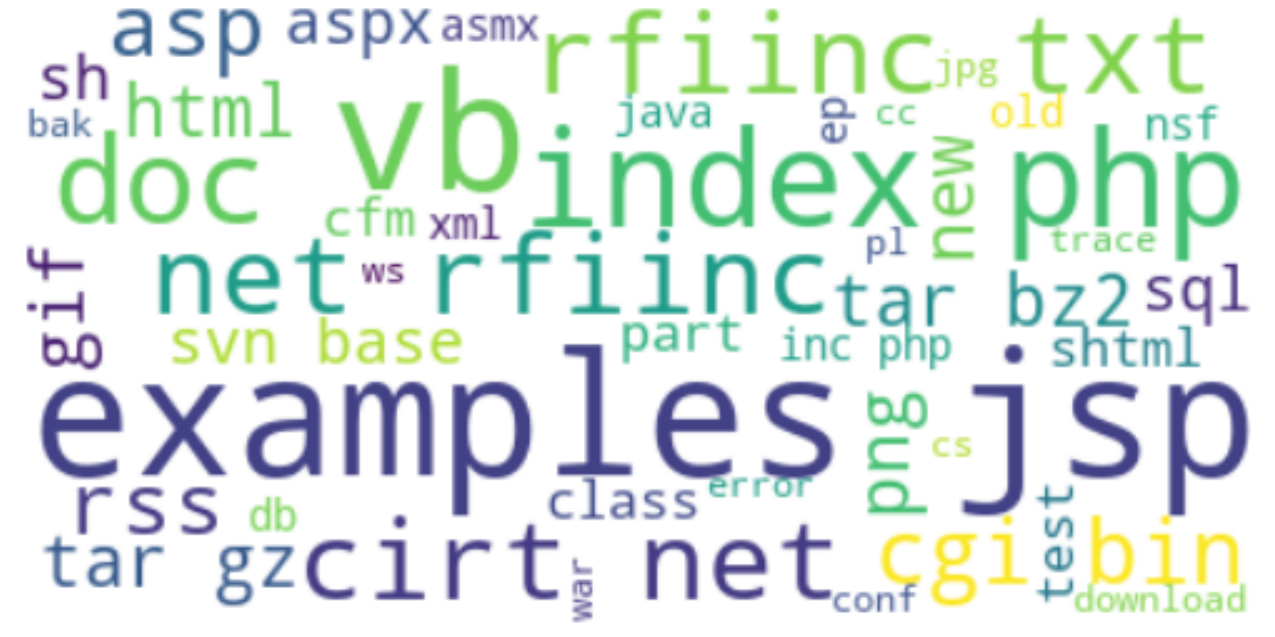}
  }
  \caption{Comparisons of payloads illustrations under WordCloud according to term frequencies: \ref{fig:badcloud}Illustration of the anomalous payloads under WordCloud, \ref{fig:goodcloud}Illustration of the normal payloads under WordCloud}
  \label{fig:wordcloud} 
\label{wordcloud}
\end{figure*}

Fig.\ref{wordcloud} illustrates the WordCloud of malicious payloads and benign payloads according to term frequency. We are easy to find that the malicious payloads look more random than a benign one and are concatenations of digits and alphabets, and thus we discuss the linguistic features in the following.

\textbf{Linguistic Features}.
To rich the feature set, we focus on linguistic features within payloads in the following. As depicted in Fig.\ref{fig:badcloud}, the number of digits offers valuable information to differentiate malicious payloads from benign payloads. The number of digits is defined as the digits occurring in a payload. The number of consecutive digits sums up the lengths of sequences of digits continuously appear. For example, Table \ref{linguisticfea} presents five linguistic features extended in our feature set. Considering a malicious payload $A$="/starnet/addons/slideshow\\ \_full.php?album\_name=288150554" and a benign one $B$= "/tests/numbertotextt\\est.php", where the number of consecutive digits are 9 and 0, respectively. The number of consecutive consonants is calculated analogously to the number of consecutive digits. The number of repeated letters are defined as the letter which occurs more than once in a payload. This feature is computed to 12 in $A$ and 4 in $B$, where the repeated letters are s, t, a, n, e, d, o, l, h, u, p, m and t, e, s, p accordingly. Similarly, the number of vowels is computed on a payload to 12 and 6 in $A$ and $B$, respectively.
\begin{table}[htbp]
\centering
\caption{Examples of linguistic features}
\begin{tabular*}{6.5cm}{lll}
\hline
\textbf{Feature} & $A$ & $B$\\
\hline
Number of Digits  & 9 &0 \\
Number of Consecutive Digits & 9 &0\\
Number of Consecutive Consonant & 19&15 \\
Number of Repeated Letters& 12 &4\\
Number of Vowels& 12 &6\\
\hline
\end{tabular*}
\label{linguisticfea}
\end{table}

\textbf{Min-Max Normalization}.
We are easy to find that the value of linguistic features is much higher than the value of TF-IDF. Min-Max Normalization is a standard method to address this issue, and it can be accomplished by rescaling the range of linguistic features to a particular range. Let $l_{i}^{j}$ be the actual value of $j-th$ linguistic feature in the $i-th$ payload. The normalized value of $l_{i}^{j}$ can be denoted as $N_{i}^{j}$. The normalization formula is given as:
\begin{equation}
N_{i}^{j}=\frac{l_{i}^{j}-{l}_{min}}{{l}_{max}-{l}_{min}}
\end{equation}
where ${l}_{min}$ and ${l}_{max}$ are the minimum, maximum value of linguistic features respectively. In this paper, we normalize the data to [0,1]. Finally, term frequency features incorporating linguistic features are concatenated into a sparse matrix representation to reduce computational complexity.

\subsubsection{Binary Classification}
According to the feature vector obtained with Section \ref{sec:feature}, it is necessary to train a classifier for determining whether a packet is benign or malicious. In this paper, we focus on supervised learning classifier with labeled samples, specifically on logistic regression since there are only two labels of payloads: benign and malicious, which is a typical binary classification problem.

Due to low computing cost and high compute speed, Logistic regression, also named as generalized linear regression, is a regression model to predict the odds of dependent variables. We choose binary logistic regression with $l2$ penalty as a classifier which has only two outputs, such as '0' and '1', which represents benign payload and malicious payload discussed in this paper, respectively. Logistic regression uses the well-known $Sigmoid$ function as below:
\begin{equation}
g\left ( z \right )=\frac{1}{1+e^{-z}}
\end{equation}
Assume that $z$ is a linear function of $x$, then $z$ can be written as:
\begin{equation}
z=\theta _{0}+\theta _{1}x_{1}+,...,+\theta_{n}x_{n}=\sum_{i=1}^{n}\theta _{i}x_{i}=\theta ^{T}x
\end{equation}
Hence, the prediction function of logistic regression can be expressed as:
\begin{equation}
h_{\theta }\left ( x \right )=g\left ( \theta ^{T} x\right )=\frac{1}{1+e^{-\theta ^{T}x}}
\end{equation}
where $h_{\theta }\left ( x \right )$ is demonstrated as the probability of the dependent variable. Given an input payload, the probability of a malicious/benign payload is given as:
\begin{equation}
P\left ( y|x;\theta  \right )=\left ( h_{\theta }\left ( x \right ) \right )^{y}\left ( 1-h_{\theta }\left ( x \right ) \right )^{1-y}
\end{equation}
For binary classification, the probability of a malicious payload is $h_{\theta }\left ( x \right )$ if $y=1$. In contrast, the probability of a normal payload is defined as $1-h_{\theta }\left ( x \right )$. Considering that training model is a time-consuming job in machine learning, hence offline training is adopted and saved to provide real-time online detection.

\subsection{Encrypted Packets Inspection}
\begin{figure}[htb]
	\centering
	\includegraphics[width=9cm]{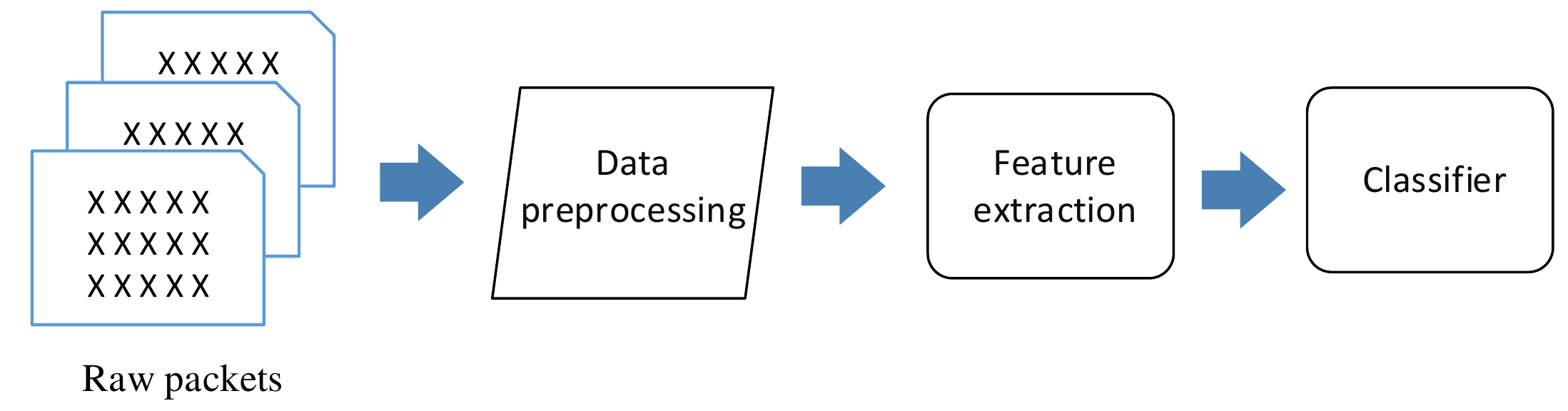}
	\caption{The workflow of detecting malicious encrypted traffic}
	\label{Fig:encryptedtraffic}	
\end{figure}

For encrypted traffic, the administrator cannot directly obtain the payload of the packet unless decrypted with the secret key. However, decrypting the encrypted traffic will affect user privacy, together with additional computational costs. In this paper, we observe that some notable features can distinguish between malicious encrypted traffic and benign traffic after analyzing an encrypted botnet dataset. Rather than decrypt network traffic, this paper proposes a statistical machine learning solution by employing some traffic features. Fig.~\ref{Fig:encryptedtraffic} shows the process of identifying malware encrypted traffic. In the following, we present some notable features in identifying malware encrypted botnet traffic.

\textbf{TLS Ciphersuites.} According to recent research from Cisco \cite{Cisco2016}, the offered TLS ciphersuites in TLS are notably distinct in malware traffic and benign traffic. Malicious traffic always uses outdated ciphersuites in \emph{clientHello} messages, while normal traffic uses more updated ciphersuites. Many new ciphersuites can only be contained after the TLS1.1 version. Therefore, the TLS version is chosen as one of the features that reflect different ciphersuites.

\textbf{TTL.} TTL specifies the maximum number of segments allowed to pass before the router drops IP packets. Generally, the number of router hops between two hosts is basically the same for a period of time. In a DDoS attack, an attacker sends a large number of malicious packets to the same host in a short period of time. As a result, there will be a large number of instantaneous data streams with the same TTL value on the client side. Thus, TTL can be a significant feature that distinguishes encrypted botnet traffic and benign traffic.

\textbf{Duration.} There is also a significant difference in the packet duration between sending benign traffic and malicious traffic. Compared to normal traffic, botnet communication has a large time interval for leaking data. In addition, the botnet will frequently send a large number of packets before the attack (5~10 pps). In this case, the duration is extremely short, which differs from normal communication.

\textbf{Ports.} Ports are often a target for attackers, such as Trojans (port 1001), worms (port 135, 445), which can be another notable feature.

\section{Evaluation}
\label{evaluation}
In this section, we first introduce the dataset utilized in this paper. Then, two binary classifiers concerning unencrypted packets inspection and encrypted packets inspection are trained offline under supervised learning. Finally, this section evaluates the performance and overheads of OFDPI.

\subsection{Dataset}
Table.~\ref{dataset1} demonstrates three datasets utilized in this paper, including two unencrypted datasets and one encrypted dataset.
\begin{table}[htb]
\setlength{\belowcaptionskip}{0.3cm}
\centering
\caption{The dataset description }
\begin{tabular}{|l|l|l|l|l|}
\hline
\textbf{Dataset}        & \textbf{Class}               & \multicolumn{2}{l|}{\textbf{Samples}} & \textbf{Total}          \\
\hline
\multirow{2}{*}{CTU-BOTNET}    & \multirow{2}{*}{encrypted}   & nomal     & 10000                     & \multirow{2}{*}{20000}  \\
\cline{3-4}
                        &                              & anomalous & 10000                     &                         \\
\hline
\multirow{2}{*}{Github Payloads} & \multirow{2}{*}{unencrypted} & normal    & 40000                     & \multirow{2}{*}{45000}  \\
\cline{3-4}
                        &                              & anomalous & 5000                      &                         \\
\hline
\multirow{2}{*}{HTTP  CSTC 2010} & \multirow{2}{*}{unencrypted} & normal    & 36000                     & \multirow{2}{*}{60000}  \\
\cline{3-4}
                        &                              & anomalous & 24000                      &                         \\
\hline
\end{tabular}
\label{dataset1}
\end{table}

\textbf{Github Payloads} \cite{github}. In order to provide a reliable machine learning model, we choose labeled datasets from Github in the offline training process, consisting of more than 40,000 benign payloads and over 5000 malicious payloads.

\textbf{HTTP CSIC 2010} \cite{CSIC}. To avoid overfitting to a particular dataset and provide a reliable machine learning model, we choose another real-world labeled dataset to evaluate the performance of OFDPI in online detection process. The HTTP CSIC 2010 is a labeled dataset that contains generated web requests with corresponding web attacks, e.g., SQL injection, cross-site scripting, buffer overflow, etc. After data preprocessing, we select 24,000 anomalous payloads and 36,000 normal payloads as our validation dataset.

\textbf{CTU-BOTNET} \cite{CTUdataset}. This encrypted dataset captures botnet traffic and normal traffic in a university network. After data preprocessing, we selected 10,000 normal flows and 10,000 botnet flows as training sets, and selected 1000 normal flows and 1000 botnet flows as test sets to ensure that the training set and the test set are mutually exclusive.

\subsection{Offline Training Test}
In this section, OFDPI trains two binary classifiers in identifying unencrypted packets and encrypted packets, respectively.

\subsubsection{Performance Metrics}
This section introduces some performance metrics when evaluating binary classifiers. Malicious packets are regarded as the positive class while negative class denotes the benign, given a set of testing packets. We select several standard classification measures to evaluate the performance when given a typical confusion matrix. We compute the following evaluation measures:

$\bm{Accuracy.}$ The detection accuracy varies with different split ratios. The higher detection accuracy, the better model performance.

$\bm{TPR=\frac{TP}{TP+FN}.}$ The True Positive Rate (TPR), also viewed as $Recall$, denotes the fraction of malicious packets among the total retrieved amount of malicious packets.

$\bm{FPR=\frac{FP}{FP+TN}.}$ The False Positive Rate (FPR), shows the fraction of benign packets classified to be malicious packets among the total retrieved amount of benign packets.

$\bm{Precision=\frac{TP}{TP+FP}.}$ Precision means the proportion of real malicious packets among all the malicious packets returned by the classifier, and it reflects there are how many real malicious packets in the classified positive results.

$\bm{F_{1}-score=\frac{2PR}{P+R}.}$ An optimal binary classifier with an expectation to have a high value of both $Precision$ and $Recall$, but they are sometimes contradictory in fact. In the analysis of the proposed binary classifier, an unbalanced problem with much more clean packets than malicious packets, we adopt $F_{1}-score$ as the harmonic average of the two values. In consideration of both $Precision$ and $Recall$, $F_{1}-score$ can well demonstrate the performance of our classifier.

\subsubsection{Unencrypted Traffic Training}
In this section, logistic regression classifier is trained under 5-fold cross validation. Note that an imbalance of data distribution exists in the dataset (white samples are far more than black samples), this paper implements \emph{stratify} method in the validation process. Besides, logistic regression is compared with other common classifiers.
\begin{figure*}[htb]
    \begin{minipage}[t]{0.5\textwidth}%
		\centering
		\includegraphics[width=2.2in]{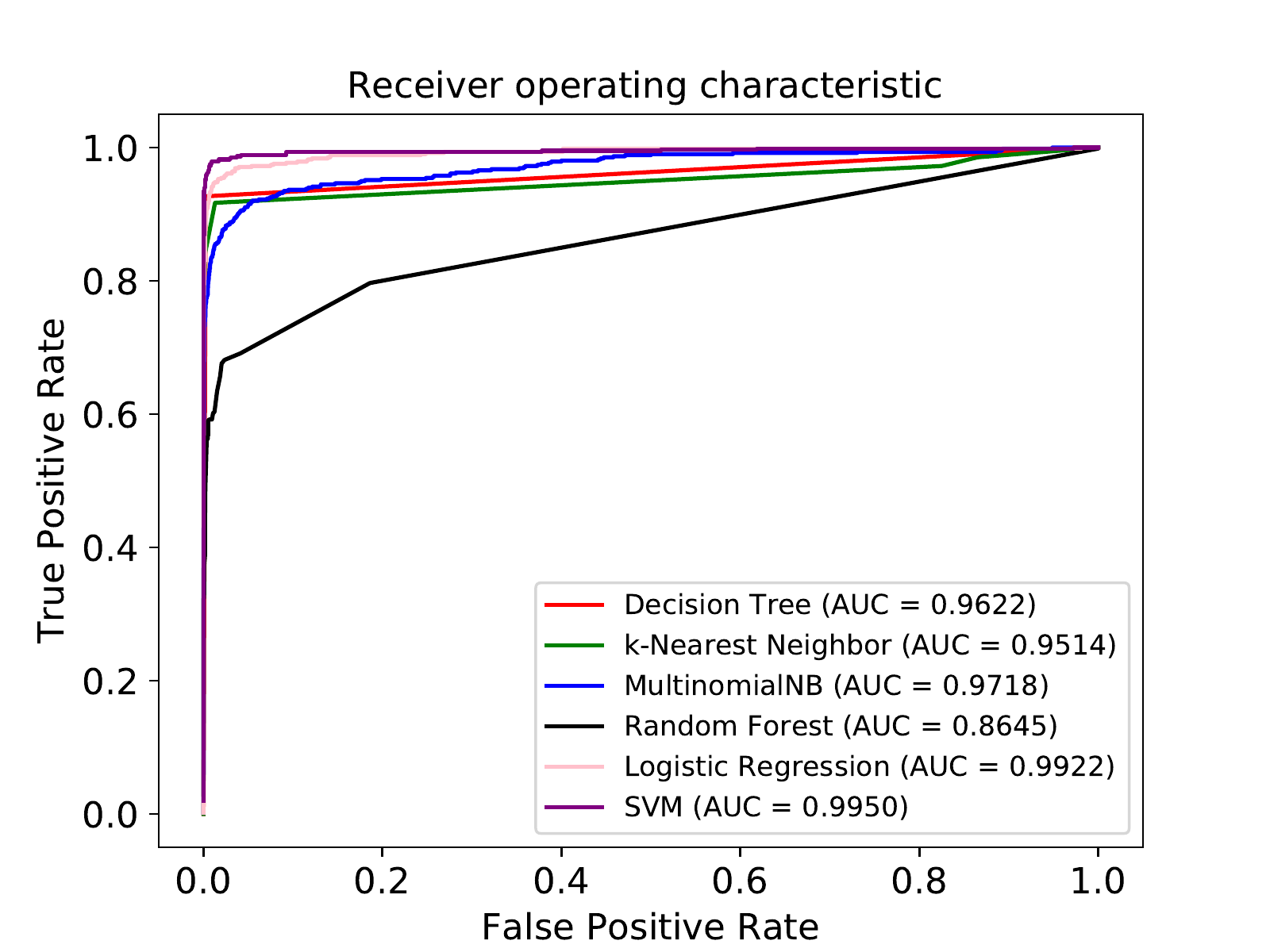}
		\caption{The Receiver operating characteristic curve of different classifiers}
		\label{roc}
	\end{minipage}
    \begin{minipage}[t]{0.5\textwidth}%
		\centering
		\includegraphics[width=2.2in]{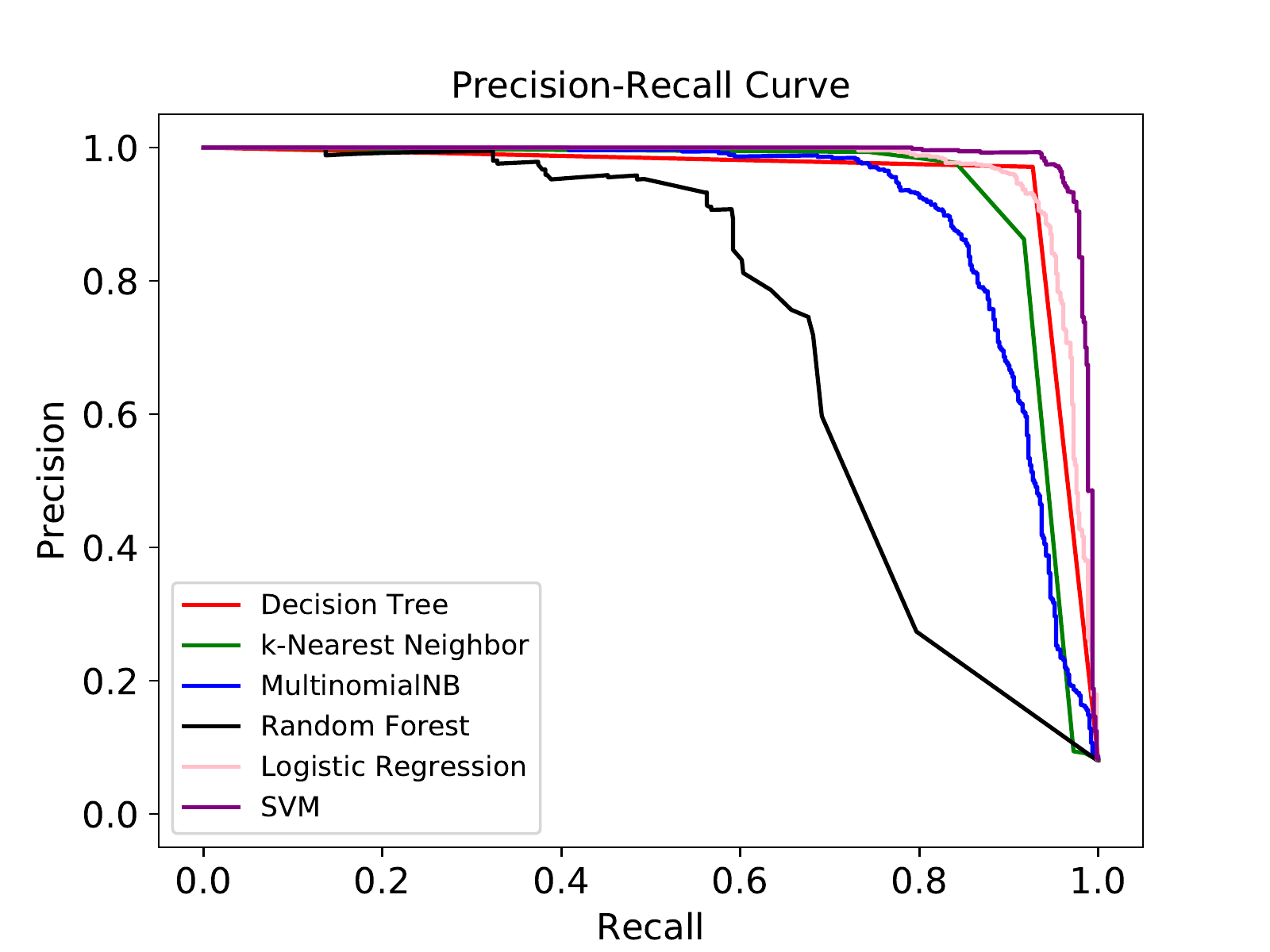}
		\caption{The Precision-Recall curve of different classifiers}
		\label{preall}
	\end{minipage}
	\begin{minipage}[t]{0.5\textwidth}
		\centering
		\includegraphics[width=2in]{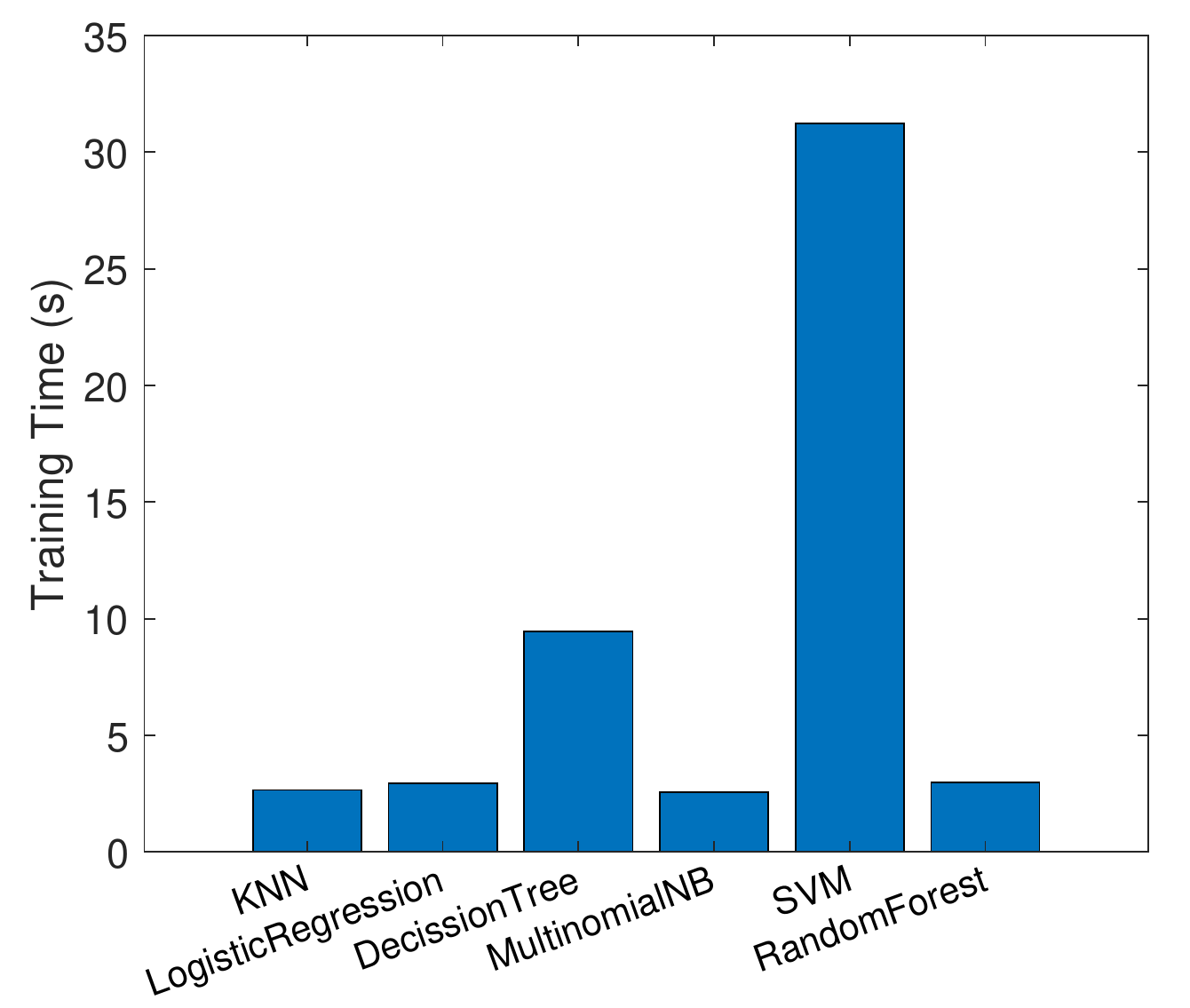}
		\caption{The training time of different classifiers}
		\label{traintime}
	\end{minipage}	
	\begin{minipage}[t]{0.5\textwidth}%
		\centering
		\includegraphics[width=2.3in]{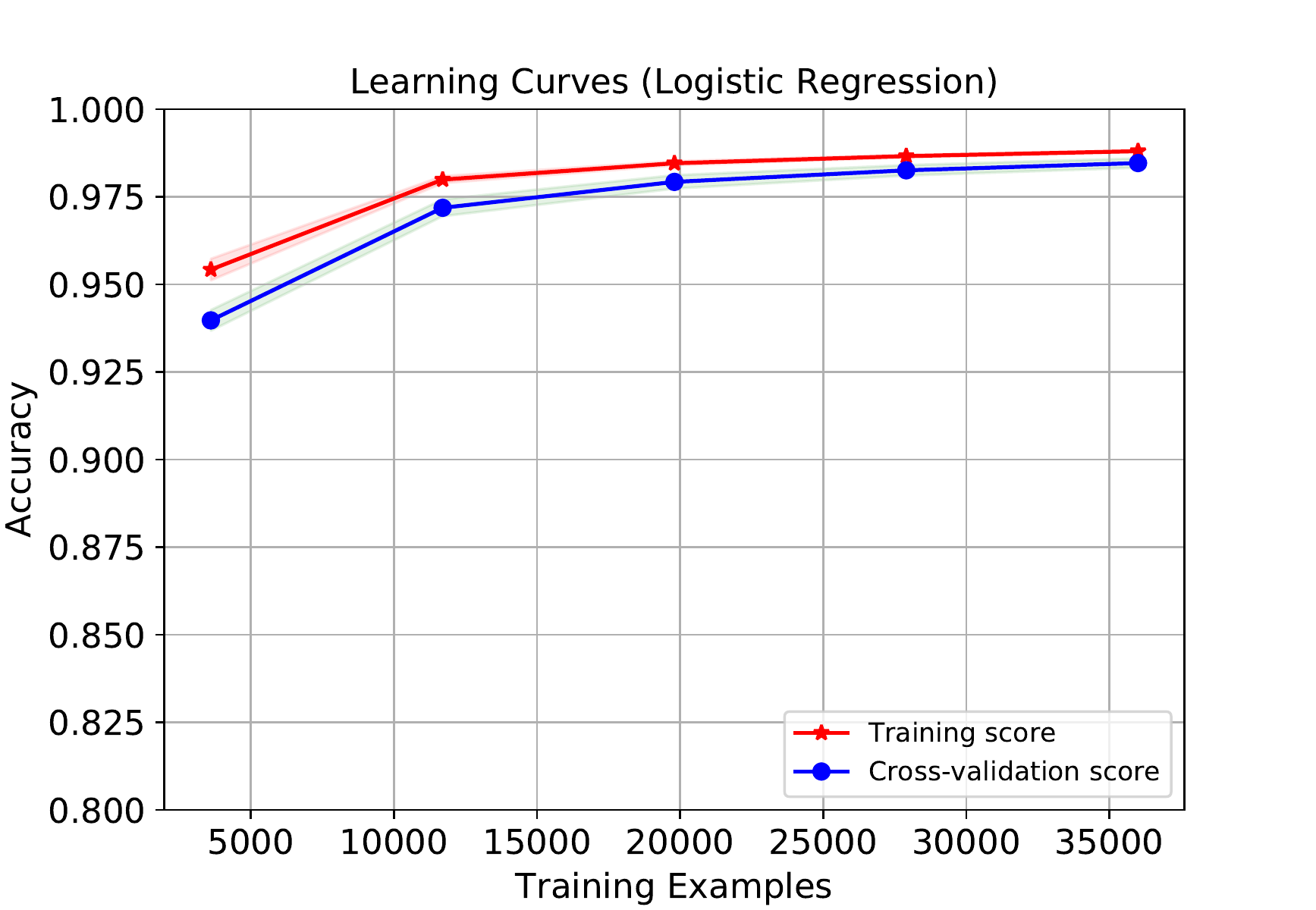}
		\caption{The learning curve of logistic regression classifier}
		\label{learningcurve}
	\end{minipage}
\end{figure*}

We adopt a Receiver Operating Characteristic (ROC) curve in the performance evaluation of binary classifiers, plotted by FPR on the horizontal axis and TPR on the vertical, which is generated with different threshold settings. Fig.~\ref{roc} presents ROC curve with six classifiers. The area under the ROC curve is always termed as AUC (Area Under Curve), which represents the probability that randomly selected benign payload is ranked higher than randomly selected malicious payload. The AUC of the logistic regression model is 99.22$\%$, which outperforms most classifiers even though it is slightly lower than that of SVM. Fig.~\ref{preall} shows the precision-recall curve of different classifiers, a graph with the recall on the x-axis and the precision on the y-axis, in which above other curves has a better performance level. In short, the precision-recall curve of SVM and logistic regression are better than other classifiers.

The training results of performance metrics are demonstrated in Table \ref{tab2}. Even though the SVM and decision tree achieve high accuracy in identifying malicious payloads, they have longer training time in comparison with logistic regression. As demonstrated in Fig.~\ref{traintime}, the training time of SVM exceeds 30 seconds which is not optimal in periodically updating models. In contrast, the average training process lasts about 2.93 seconds under logistic regression, with an accuracy of 98.96$\%$. In summary, logistic regression is better than other classifiers in identifying anomalous payloads.
\begin{table}[htbp]
\caption{Comparisons of different classifiers}
\begin{center}
\begin{tabular}{|c|c|c|c|c|}
\hline
\textbf{Different}&\multicolumn{4}{|c|}{\textbf{Metrics(\%)}} \\
\cline{2-5}
 \textbf{Classifiers}& \textbf{\textit{Accuracy}}& \textbf{\textit{Precision}}& \textbf{\textit{Recall}}& \textbf{\textit{F1-score}} \\
\hline
KNN& 0.9741&	0.9349&	0.865&	0.8986\\
\hline
Logistic Regression& 0.9896&0.9732&	0.91&	0.9486\\
\hline
Decision Tree&0.9864&	0.9406&	0.9284&	0.9345 \\
\hline
MultinomialNB &0.9768&	0.9563&	0.7479&	0.8394\\
\hline
Random Forest&0.9219&	0.9996&	0.1505&	0.2524\\
\hline
SVM &0.9926 &0.9721 &0.9349 &0.9536\\
\hline
\end{tabular}
\end{center}
\label{tab2}
\end{table}

In order to get the learning curves of logistic regression classifier, we implement 5-fold cross validation with 50 iterations that the validation set is randomly selected with 20$\%$. Fig.~\ref{learningcurve} shows the learning curve of a binary logistic regression classifier. It can be easily seen that the training score is higher than 0.95 at the beginning and it increases slowly. The validation score increases with more training examples, and the average score is more than 0.925, but a minute difference still exists between the training accuracy and validation accuracy, which indicates slight overfitting with the model. We can quickly conclude that adding more training examples is likely to help better the model with current logistic regression algorithm.

Even though OFDPI provides deep packet inspection with a high detection accuracy of 98.96$\%$, it heavily relies on traditional feature engineering. In particular, we extract the payload features including tri-gram frequency-based TF-IDF and linguistic features. However, the feature extraction is a complicated and time-consuming process. In contrast, \cite{CNN2018} provides an end-to-end payload classification without feature extraction using deep learning models. They adopt a word embedding technique to encode raw payloads and then apply the CNN-based and RNN-based classification approach to learn features from initial payloads. They achieve a high detection accuracy of 99.36$\%$ over the DARPA1998 dataset. Unfortunately, encrypted payloads are not validated in \cite{CNN2018}.

In this paper, OFDPI performs traffic classification under supervised learning. Both unencrypted payloads and encrypted payloads are trained with labeled dataset whereas the labeled data is rare and expensive in realistic scenarios. Recently, \cite{Changhe2018} achieves a rather high detection accuracy with 60$\%$ labeled data under semi-supervised machine learning.  Even the performance of the SDN controller is a lack of evaluation in \cite{Changhe2018}, semi-supervised learning is still a promising direction in deep packet inspection in SDN. This can be discussed in our future work.

\subsubsection{Encrypted Traffic Training}
This section adopts an encrypted botnet dataset from CTU \cite{CTUdataset} to train a binary classifier with some machine learning algorithms. Table.~\ref{encryptedresults} shows the training results of different classifiers.

It is easy to observe that SVM has the worst classification performance even if its recall value is high. In comparison, Decision tree achieves the highest accuracy in detecting malware encrypted botnet traffic with 99.15$\%$. This high value is partial because the features extracted from the CTU dataset are distinguished. Thus, OFDPI chooses a decision tree model to detect encrypted malware traffic.
\begin{table}[htbp]
\caption{Training results of different classifiers with encrypted traffic}
\begin{center}
\begin{tabular}{|c|c|c|c|c|}
\hline
\textbf{Different}&\multicolumn{4}{|c|}{\textbf{Metrics(\%)}} \\
\cline{2-5}
 \textbf{Classifiers}& \textbf{\textit{Accuracy}}& \textbf{\textit{Precision}}& \textbf{\textit{Recall}}& \textbf{\textit{F1-score}} \\
\hline
SVM& 0.5635&	0.533903&	0.98&	0.696136\\
\hline
MLP& 0.968&0.947419&	0.991&	0.968719\\
\hline
Decision Tree&0.9915&	0.997974&	0.985&	0.991444 \\
\hline
\end{tabular}
\end{center}
\label{encryptedresults}
\end{table}

This paper also compares the performance with nDPI, a popular OpenDPI library used for deep packet inspection. Table ~\ref{nDPIbot} shows the results when nDPI processes the malware botnet dataset from CTU. The results show that all the traffic is divided into several security levels, including safe, acceptable, unsafe, fun and unrated. nDPI classifies traffic with finer granularity in comparison with OFDPI. Nevertheless, 12.6085$\%$ of the tested malware traffic is misjudged as normal, and most traffic is considered acceptable and unrated.
\begin{table}[htb]
\setlength{\belowcaptionskip}{0.3cm}
\caption{Detection results with nDPI}
\centering
\begin{tabular}{l|l|l|l|l|l}
\hline
\textbf{Security Level} & safe    & acceptable & fun     & unsafe & unrated  \\
\hline
\textbf{~ ~ percent}    & 12.6085 & 39.0352    & 0.29043 & 0.0791 & 47.9868  \\
\hline
\end{tabular}
\label{nDPIbot}
\end{table}

\subsection{Online Detection}
This section presents the online detection with OFDPI. In particular, we adopt a famous IP blacklist to perform early detection, and then sample traffic according to the LP-based traffic sampling window.

\subsubsection{Evaluation Testbed}
This paper implements a prototype of OFDPI by using a Mininet network emulator and an SDN controller running on a virtual machine with Intel Core i7-6700 processor. The SDN controller in OFDPI needs to receive notifications from the DPI engine and then sends the \texttt{FLOW\_MOD} message if an anomalous packet is detected. As a result, the deep packet inspection process inevitably induces the network latency. The lower latency requirement leads to our preference for Ryu SDN controller since Ryu exhibits the best latency results in comparison with ONOS, OpenDayLight and Floodlight \cite{Lusani}. In addition, Ryu has better TCP throughput as compared to POX \cite{Ali2018}. Moreover, Ryu is python-based which is much easier to integrate with existing mainstream machine learning libraries (e.g., scikit-learn in this paper).

\begin{figure}[htb]
	\centering
	\includegraphics[width=8cm]{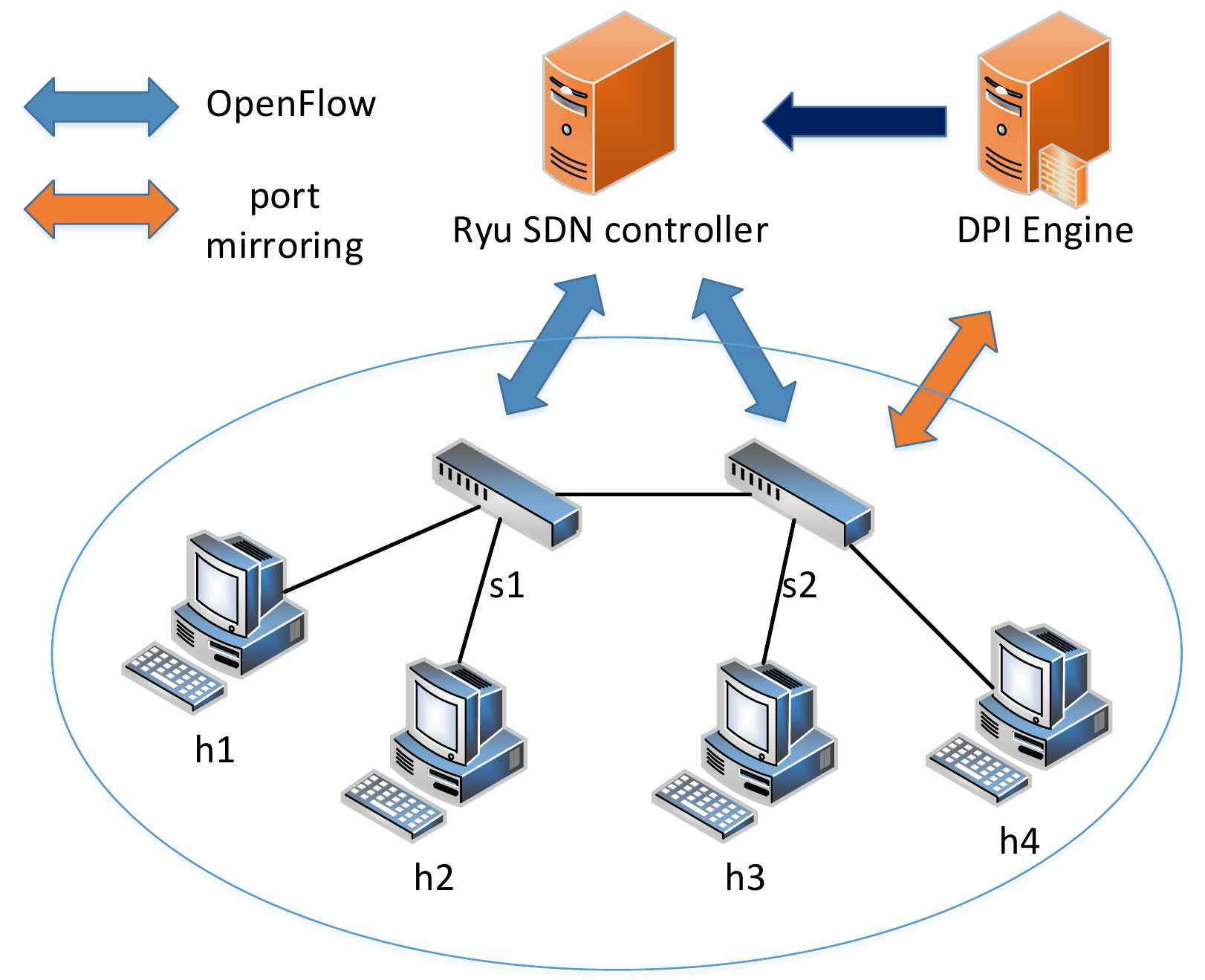}
	\caption{The network topology}
	\label{Fig:topo}	
\end{figure}

The network topology is shown in Fig.~\ref{Fig:topo}. The controller is connected to two switches, each of which is connected to two hosts. The maximum bandwidth of each link is set to 10 Mbps. The OFDPI engine is placed on another server to ease the burden on the controller. When h1 attempts to connect to h4, traffic traversing the switch will be redirected to the DPI engine through port mirroring.

\subsubsection{Detection with New Dataset}
\label{detection}

In our experiments, a well-known IP blacklist from Cisco \cite{Blacklist} is applied in the SDN controller to perform early IP address filtering. Initially, TCP packets containing the payloads reach the OpenvSwitch, they invoke datapath module in the kernel and then extract key values (e.g., IP addresses and MAC addresses) to match against flow entries in the flow table. The IP blacklist stored in the Ryu SDN controller is used to filter anomalous IP addresses. In our experiments, OFDPI only blocks a few packets containing the malicious IP address, whereas a large number of packets successfully match against the flow entries.
\begin{figure}[htb]
	\centering
	\includegraphics[width=7.5cm]{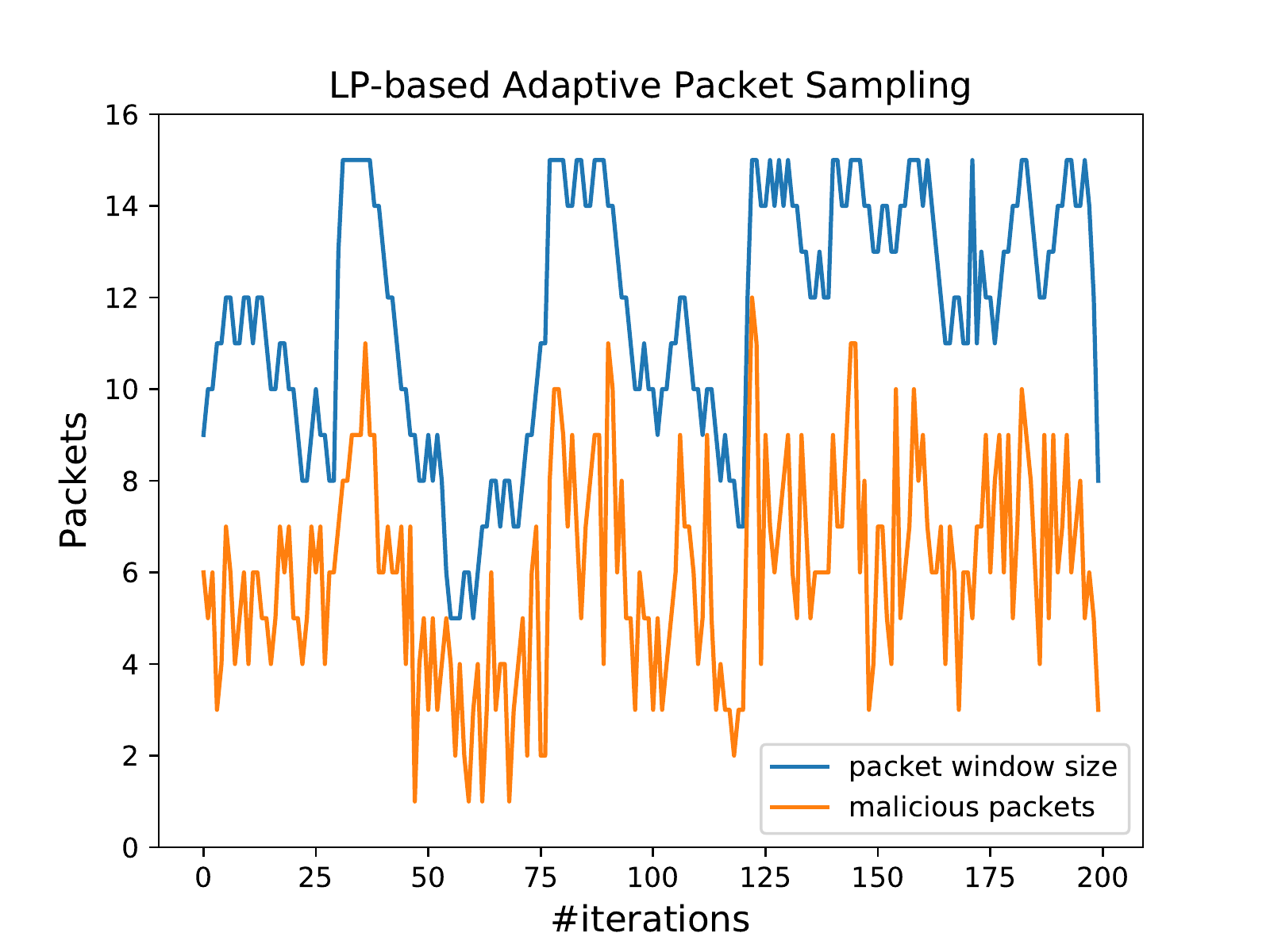}
	\caption{LP-based adaptive packet sampling results}
	\label{Fig:sampresult}	
\end{figure}

Then, OFDPI applies an adaptive packet sampling to those packets that bypass the IP blacklist. Unreasonable sampling rate may cause a burden on the CPU \cite{sampling2018}. Considering the throughput of each link, $m$ is determined as 100 in our scenarios, i.e., we select a packet window size from every 100 packets. The maximum and minimum packet window is set to be 5 and 15, respectively. Fig.~\ref{Fig:sampresult} illustrates the detection results under linear prediction model. Each packet window is adaptively selected according to previous ten samples (i.e., let $N=10$) since the current traffic pattern has little to do with a large number of previous samples. The yellow graph in Fig.~\ref{Fig:sampresult} shows the detected malicious packets, and the blue one deals with the packet window size. It shows that the number of malicious packets is linearly related to the packet window size. When the packet window size fluctuates around the maximum packet window size, it usually achieves the best detection results with more malicious packets but at the same time with an increase of resource consumption (detailed discussion in Section~\ref{RTToverheads}). However, the limitations in the LP-based packet sampling mechanism are obvious. We only utilize a simple adjustment algorithm to expand or contract the packet window size. As suggested in \cite{Zhang2013An}, more accurate adjustment algorithms should be added into the LP-based mechanism, especially the latest network attacks are more uncertain and dynamic \cite{Hoque2014}.

\begin{figure}[htb]
  \centering
  \includegraphics[width=7cm]{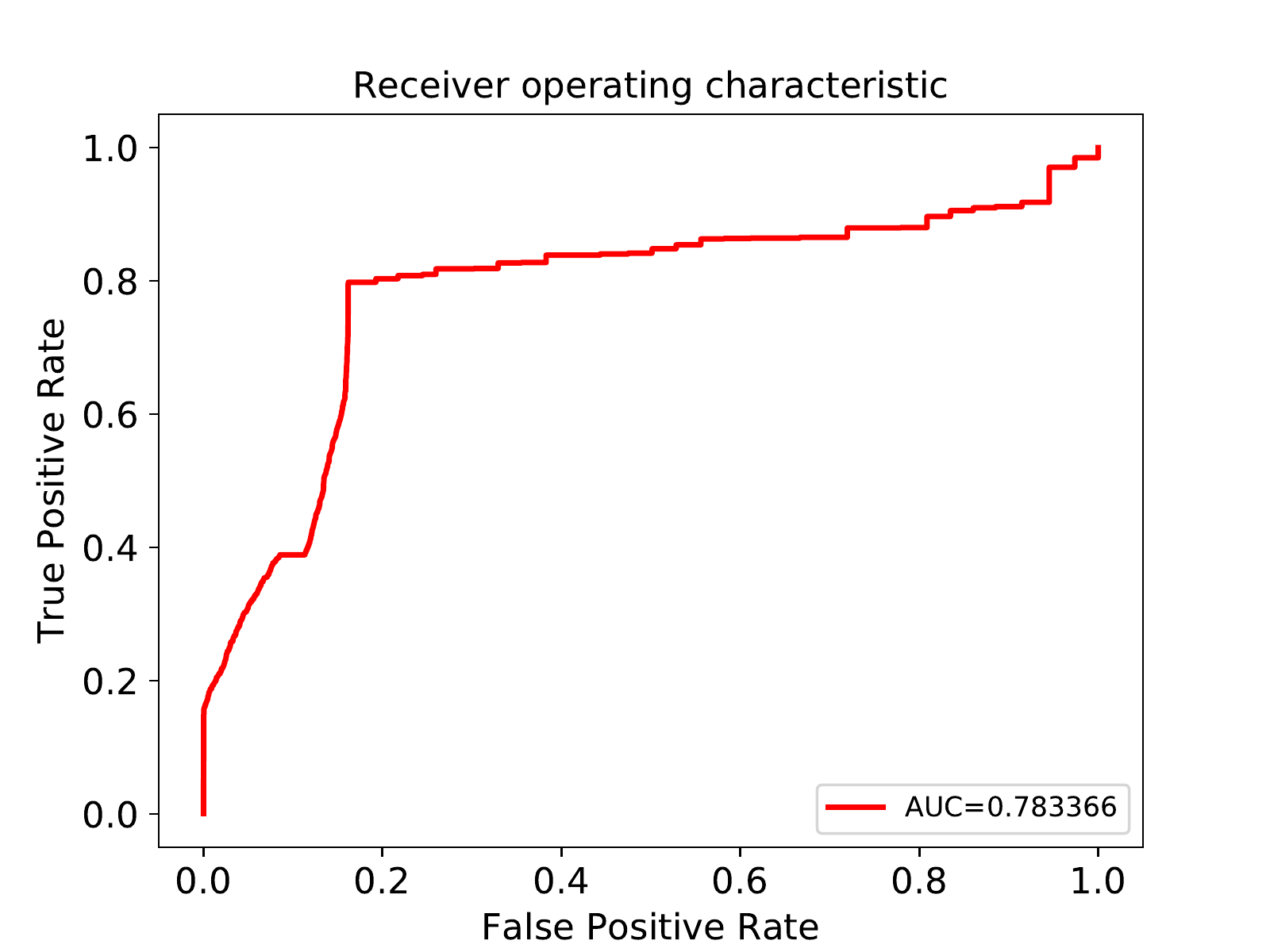}\\
  \caption{The Receiver operating characteristic of dataset CSIC-2010}
  \label{Fig:roc2010}
\end{figure}

\begin{figure}[htb]
  \centering
  \includegraphics[width=7cm]{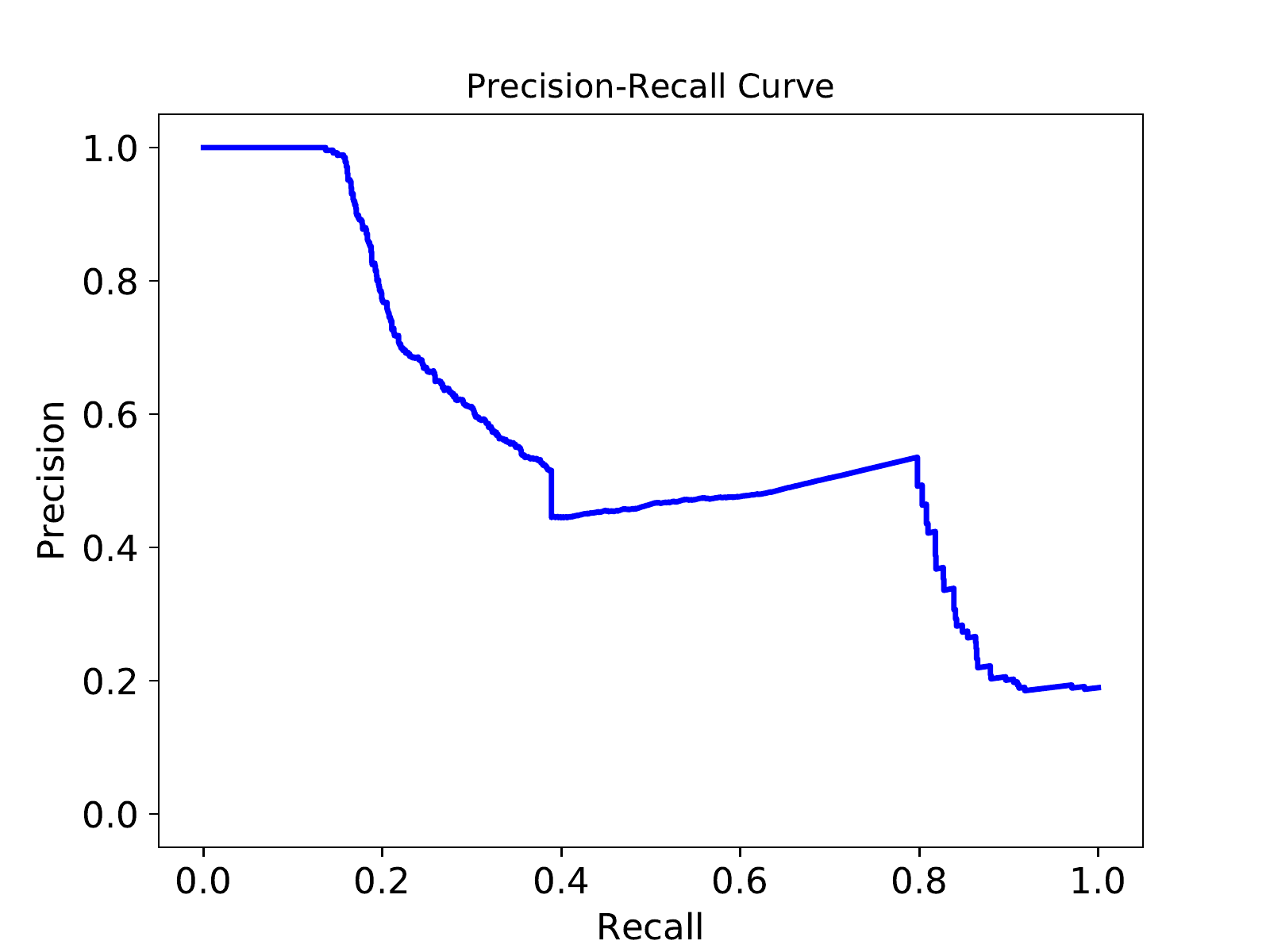}\\
  \caption{The Precision-Recall curve of dataset CSIC-2010}
    \label{Fig:precall}
\end{figure}

As discussed above, it is likely to overfit with machine learning algorithms to a particular dataset especially where an imbalance of data distribution exists. In addition to address the class imbalance issue of a dataset in the training process (e.g., focal loss and stratify), we collect a supernumerary real dataset to validate the detection accuracy to avoid overfitting \cite{CSIC}. Experimental results reveal that the detection accuracy of OFDPI is 82.31$\%$ with the new dataset, and the ROC curve is as demonstrated in Fig.~\ref{Fig:roc2010}, in which the AUC under new dataset is 78.33$\%$. The precision-recall curve is demonstrated in Fig.~\ref{Fig:precall} with a recall of 56.51$\%$. The results under a distinct dataset are not satisfied enough which needs further improvement on feature extraction.

\subsection{Performance Evaluation}
\label{RTToverheads}
This section evaluates the throughput of the SDN controller when identifying packets with OFDPI. In addition, CPU and memory utilization are also presented in OFDPI.

\subsubsection{Throughput of SDN Controller}
\begin{figure}[htb]
  \centering
  \includegraphics[width=7cm]{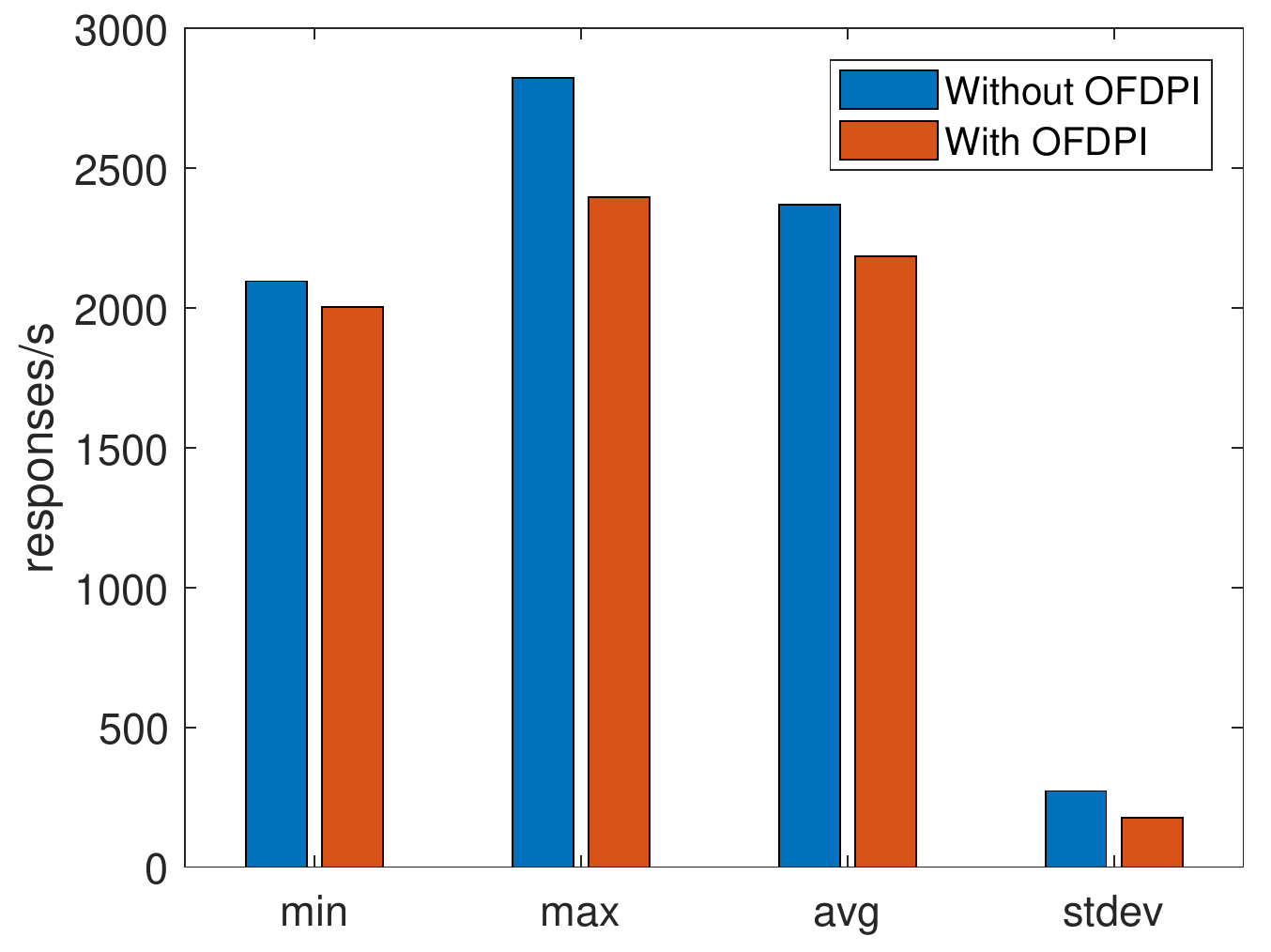}\\
  \caption{The throughput of the Ryu SDN controller}
  \label{throughputcompare}
\end{figure}

Adding additional monitoring functions to the SDN controller may introduce some performance bottlenecks to the controller itself. This section uses the Cbench \cite{cbench} tool to test the throughput of the SDN controller with the OFDPI by continuously looping through the new Packet$\_$in messages. Initially, an SDN-based network is constructed as shown in Fig.~\ref{Fig:topo}, in which h1 persistently sends packets to the h4. Then, Cbench connects to the Ryu controller by emulating 14 switches, and each switch is associated with 1000 unique MAC addresses. We test 5 loops and each loop lasts for 10,000 ms, ignoring the first "warmup" and "cooldown" loops. Finally, we record the minimum, maximum, average and standard deviation values of the throughput of the Ryu SDN controller. In comparison, we also test the throughput of the Ryu SDN controller without OFDPI. The comparison results are demonstrated in Fig.~\ref{throughputcompare}.

Fig.~\ref{throughputcompare} shows that the controller is able to process an average of 2370.35 flows per second without OFDPI. While this value is reduced to 2185.36 responses/s with OFDPI. It is clear that OFDPI reduces the controller's ability to process flows, but this slight performance degradation is acceptable.

\begin{figure*}[htbp]
\centering
\begin{minipage}[t]{0.48\textwidth}
\centering
\includegraphics[width=2.5in]{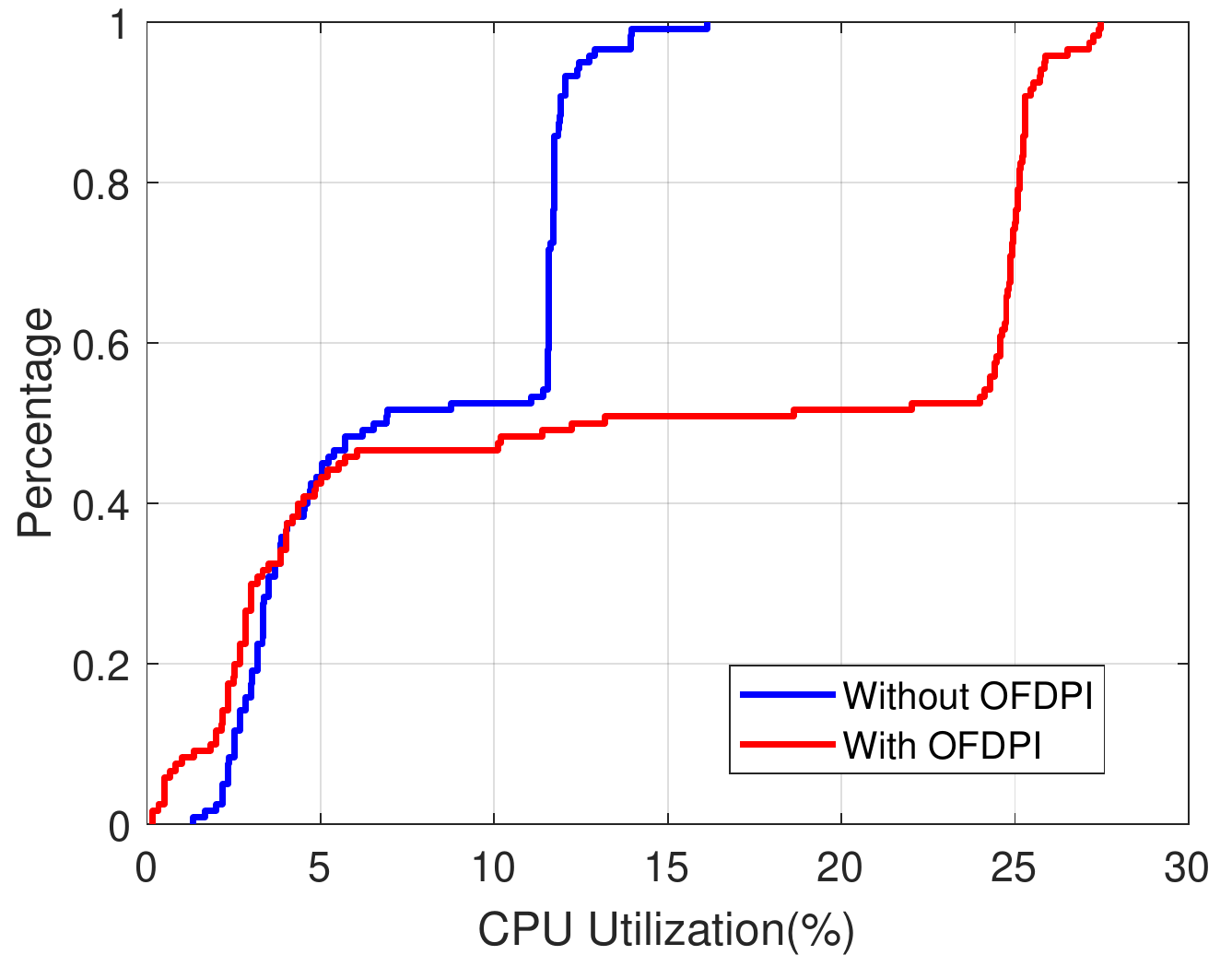}
\caption{The probability distribution of the CPU utilization}
\label{Fig:cpu}
\end{minipage}
\begin{minipage}[t]{0.48\textwidth}
\centering
\includegraphics[width=2.5in]{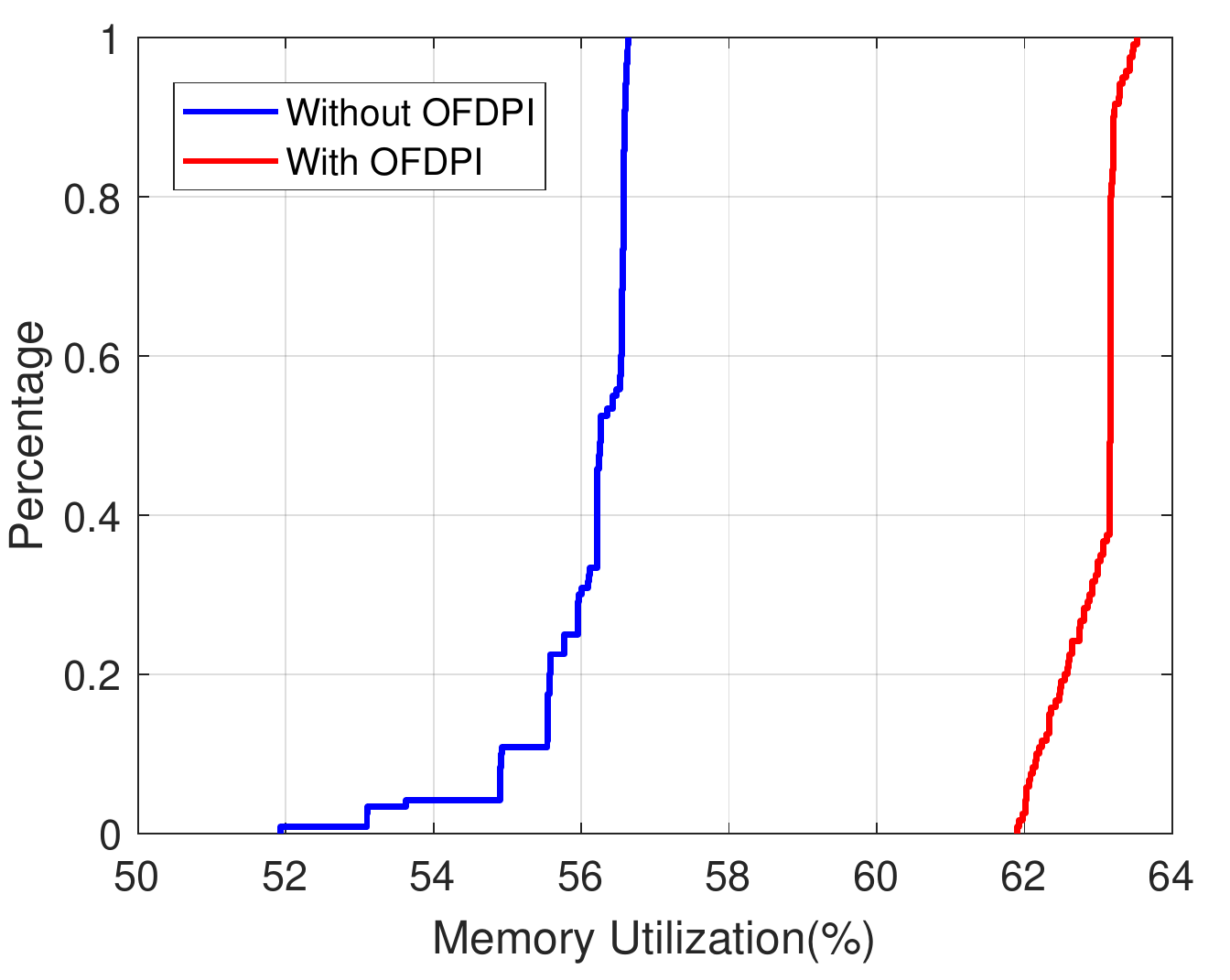}
\caption{The probability distribution of the memory utilization}
\label{Fig:mem}
\end{minipage}
\end{figure*}
\subsubsection{Overheads}
Besides, we evaluate the overheads caused by OFDPI concerning CPU and memory. In the first experiment, an SDN-based network with Mininet platform and the Ryu SDN controller is constructed, in which the $OFDPI\_reply\_handler$ thread is monitoring the underlying traffic. When the data set \textbf{HTTP CSTC 2010} is running on the forwarding plane as section ~\ref{detection}, CPU and memory utilization of the current system is recorded per second. Finally, we get a total of 120 CPU and memory records, respectively. In comparison, the CPU and memory utilization without OFDPI are also tested under the same scenario.

Figure~\ref{Fig:cpu} illustrates the cumulative distribution function (CDF) plot of the CPU utilization in different experimental settings. There is one bearing at each graph where the blue line represents the unmodified SDN controller and the red line with OFDPI. The data are distributed around values 12 and 25, respectively. Obviously, the average CPU utilization with OFDPI is greater than that without OFDPI. It is partly because the Ryu SDN controller consumes much more CPU when monitoring the underlying traffic with OFDPI.

Figure~\ref{Fig:mem} shows a CDF plot of the memory utilization of different settings. 30$\%$ of the values in the distribution are less than or equal to the 56$\%$ memory utilization without OFDPI. In contrary, the memory utilization caused by OFDPI is greater with an average value of 63$\%$. In summary, the slight increase in overheads caused by OFDPI is acceptable in our experiments. In comparison, the overhead of the SDN controller in OFDPI is higher than that of \cite{lin2015}. In [R2], only a few percents of input traffic is sent to the controller, since they extend the data plane to redirect traffic to the DPI module or other NFV modules by service function chaining which eases the burden of the controller.  However, the detection accuracy of the DPI module is not mentioned in \cite{lin2015}.

\section{Conclusion}
\label{conclusion}
Deep packet inspection (DPI) in software-defined networking (SDN) remains restrictions in the presence of a large volume of data. Despite third-party DPI tools, this paper proposes a novel OpenFlow-enabled deep packet inspection (OFDPI) approach in SDN using machine learning algorithms. OFDPI enables deep packet inspection for both unencrypted traffic and encrypted traffic by training two binary classifiers respectively. In addition, OFDPI is able to sample suspicious packets with a packet window based on linear prediction. We evaluate the performance of OFDPI on the Ryu SDN controller and the Mininet platform with real-world datasets. OFDPI achieves a rather high detection accuracy for both encrypted traffic and unencrypted traffic, as well as acceptable overheads.


\bibliographystyle{IEEEtran}

\begin{IEEEbiographynophoto}{Qiumei Cheng} is currently pursuing the Ph.D. degree with the College of Computer Science and Technology, Zhejiang University. Her research interests include software-defined network security, intrusion response system, traffic monitoring and reinforcement learning.
\end{IEEEbiographynophoto}
\vspace{-0.4cm}
\begin{IEEEbiographynophoto}{Chunming Wu} received the Ph.D. degree in computer science from Zhejiang University in 1995. He is currently a Professor with the College of Computer Science and Technology, Zhejiang University. His research fields include software-defined networks, reconfigurable networks, proactive network defense, network security, network virtualization, the architecture of next-generation Internet, and intelligent networks.
\end{IEEEbiographynophoto}
\vspace{-0.4cm}
\begin{IEEEbiographynophoto}{Haifeng Zhou} received the Ph.D. degree in computer science and technology from Zhejiang University in 2018. He is currently a researcher with the College of Control Science and Engineering, Zhejiang University. His research interests include software-defined network security, proactive network defense, intelligent networks and security systems, cloud security.
\end{IEEEbiographynophoto}
\vspace{-0.4cm}
\begin{IEEEbiographynophoto}{Dezhang Kong}  received the B.E degree in information security from the Huazhong University of Science and Technology in 2018. He is currently pursuing the Ph.D. degree with the College of Computer Science and Technology, Zhejiang University. His research interests include AI security, network security and software-defined network security.
\end{IEEEbiographynophoto}
\vspace{-0.4cm}
\begin{IEEEbiographynophoto}{Dong Zhang}  received the Ph.D. degree in computer science from Zhejiang University in 2010, and currently is an associate professor of College of Mathematics and Computer Science at Fuzhou University. His research areas include software defined networking, network virtualization and Internet QoS.
\end{IEEEbiographynophoto}
\vspace{-0.4cm}
\begin{IEEEbiographynophoto}{Junchi Xing} received the Ph.D. degree with the College of Computer Science and Technology, Zhejiang University, Hangzhou, China. His research interests include software-defined networks, software-defined network security, and cloud security.
\end{IEEEbiographynophoto}
\vspace{-0.4cm}
\begin{IEEEbiographynophoto}{Wei Ruan} received M.S. and Ph.D. degrees from Zhejiang University majoring in the Department of Energy in 1997 and 2000, respectively. He has joined the SUPCON Group Co.,Ltd since 2000 and has served as major project director, deputy chief engineer, and vice president of Zhejiang Supcon Research Co.,Ltd. until 2016. He has been engaged in the research on the software and hardware of the national strategic equipment automatic control system, the optimization control strategy, and the field engineering application over a long period of time.
\end{IEEEbiographynophoto}

\end{document}